# Assessing the climate benefits of afforestation: processes, methods, and frameworks

Kevin Bradley Dsouza[a*], Enoch Ofosu[a], Jack Salkeld[b], Richard Boudreault[c,a], Juan Moreno-Cruz[d*], Yuri Leonenko[a,b*]

a - Department of Earth and Environmental Sciences, University of Waterloo
b - Department of Geography and Environmental Management, University of Waterloo c - Techaero
d - School of Environment, Enterprise and Development, University of Waterloo
* - Corresponding authors

## Abstract
Afforestation greatly influences several earth system processes, making it essential to understand these effects to accurately assess its potential for climate change mitigation. Although our understanding of forest-climate system interactions has improved, significant knowledge gaps remain, preventing definitive assessments of afforestation's net climate benefits. In this review, focusing on the Canadian northern boreal and southern arctic, we identify these gaps and synthesize existing knowledge. The review highlights regional realities, Earth's climatic history, uncertainties in biogeochemical (BGC) and biogeophysical (BGP) changes following afforestation, and limitations in current assessment methodologies, emphasizing the need to reconcile these uncertainties before drawing firm conclusions about the climate benefits of afforestation. Finally, we propose an assessment framework which considers multiple forcing components, temporal analysis, future climatic contexts, and implementation details. We hope that the research gaps and assessment framework discussed in this review inform afforestation policy in Canada and other circumpolar nations.

## Introduction
Climate change poses a critical threat to humanity, with observed and projected warming rates unprecedented in the current interglacial period. Unless we act swiftly to reduce greenhouse gas (GHG) emissions and begin sequestering existing accumulated atmospheric GHGs, climate change impacts will likely intensify in the coming years, impacting ecosystems worldwide [1]. Some ecosystems are more vulnerable than others, with high-latitude ecosystems such as  the boreal and Arctic warming two to four times faster than the global average [2, 3], making them highly sensitive areas needing stewardship. Canada is home to one-third of the boreal biome that envelops the global northern hemisphere, which is a significant store of terrestrial carbon [4], with managed boreal forests alone storing ~28 gigatonnes (Gt) of carbon [4].

The Intergovernmental Panel on Climate Change (IPCC) recognizes the vast potential of forests to sequester carbon dioxide ($CO_2$) [1]. Afforestation is projected to provide substantial sequestration benefits this century, estimated at ~4.9 $GtCO_2$/year globally [5]. The Canadian government's "Two Billion Trees" program [6] exemplifies the significant interest in afforestation, particularly in the boreal region [7]. However, it is essential to consider that forests impact the climate in complex ways, extending beyond carbon sequestration to influence albedo, surface energy balance, hydrological cycles, and permafrost dynamics. While significant progress has been made in understanding the impacts of forests on regional dynamics and global climate processes, many knowledge gaps remain, hindering the consideration of these effects in existing assessments of afforestation's climate benefits [8-10].

In this work, we explore the interconnections of forest processes (see Box 1a), revealing that afforestation is a more complex decision than it initially appears. We explore the unique realities of the northern boreal and southern arctic regions (see Box 1b for land cover map), including permafrost, hydrology, snow behavior, and general forest considerations such as non-radiative processes, soil carbon, forest structure, and chemical emissions (see Box 1a). Additionally, we examine what can be learned from forest behavior during Earth's climatic history and the uncertainties in forest dynamics under projected climate change this century. We also highlight the need to reconcile remote sensing-based methodology with climate models and point out the methodological limitations of existing afforestation assessments. Finally, we discuss how these insights can be used to improve afforestation project modeling and outline a path forward for analysis, planning, and policy-making. For a quick introduction to the acronyms and abbreviations used in this review refer to Table 1 in the supplementary.

a

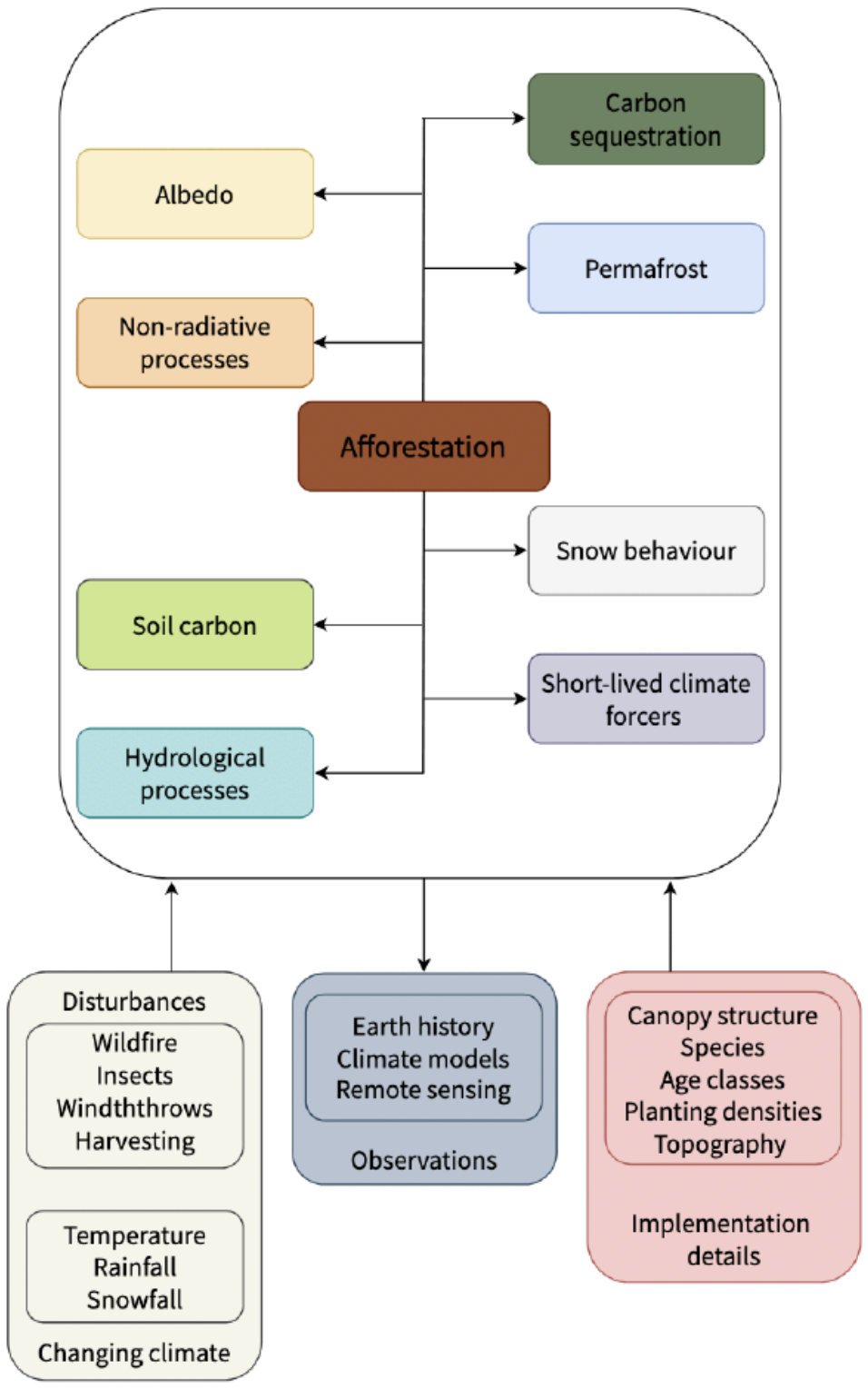

b

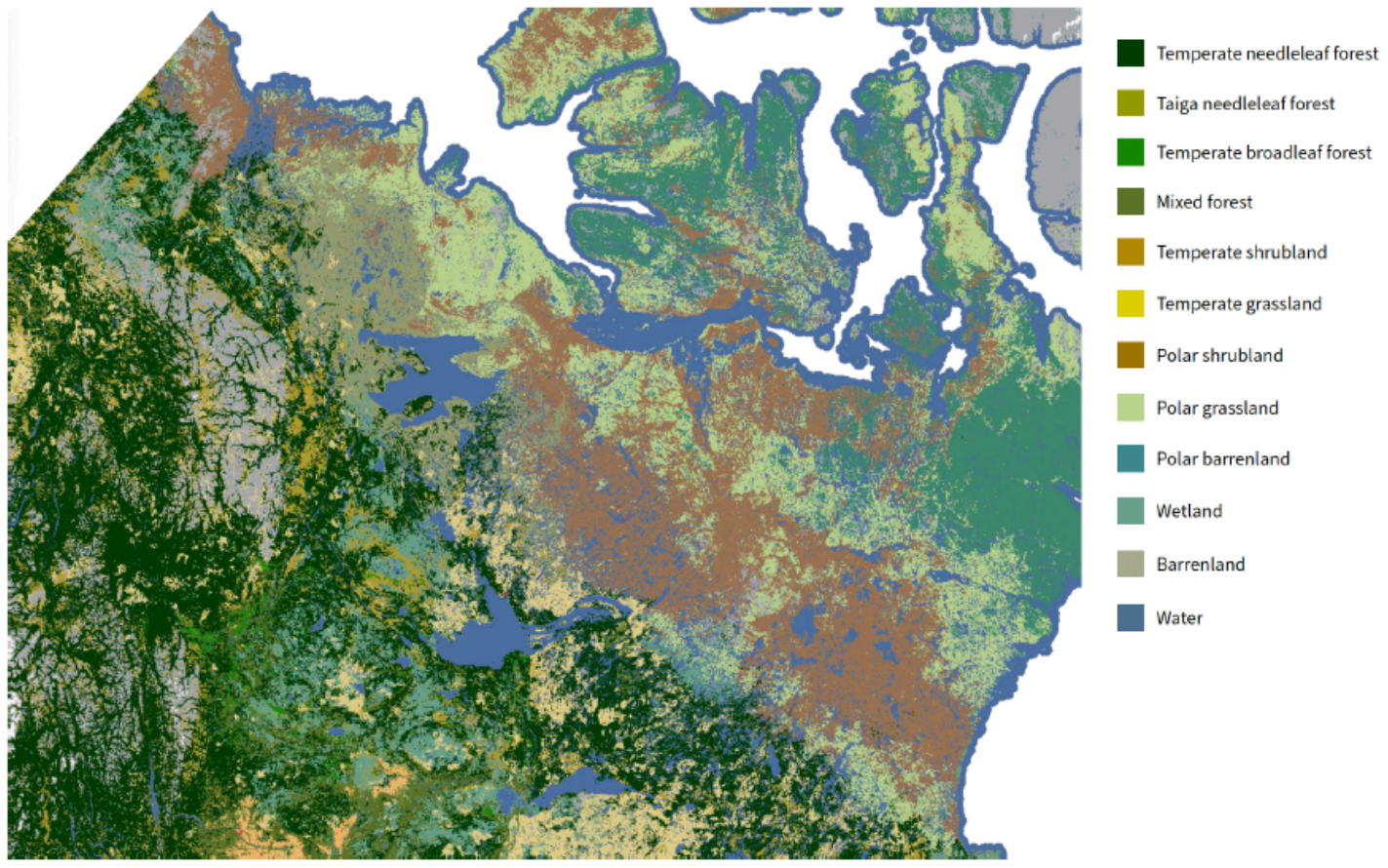

Box 1: a) A graph showing various processes that afforestation influences, observations used to study these processes, and implementation details that are crucial. Apart from BGC processes such as carbon sequestration and emission of short-lived climate forcers (SLCFs), afforestation influences a variety of other processes including albedo (radiative), non-radiative and hydrological processes, and dynamics such as permafrost, snow, and soil organic carbon (SOC) dynamics. The influence that afforestation has on these processes and dynamics can be studied using observations from remote sensing, climate model simulations, and Earth's geologic history. A changing climate is expected to affect afforestation and all its interlinked processes by altering disturbance regimes like wildfire and insects, as well as modifying climate variables such as temperature and precipitation. Implementation details including the group of species chosen to be afforested, age distributions in a given forested area, and planting densities, changes canopy structure and affects various processes linked to afforestation. Moreover, the topography chosen for afforestation affects the overall surface energy balance by altering solar illumination, snow behavior, and hydrology. b) Land cover classification for north-western boreal [11]. The northwestern-boreal is predominantly made of temperate and taiga needleleaf (evergreen) forests, but also has a significant percentage of temperate broadleaf (deciduous) and mixed forests. In order to determine which species would grow best in the gaps between forested regions and surrounding non-forested regions, the tradeoffs between different components in a) need to be considered.

## Regional realities and processes

Each ecosystem has unique characteristics and key drivers that play a crucial role in its functioning and sets it apart from other ecosystems. In the subsequent sections, we expand upon these critical processes and realities central to the boreal and southern arctic regions.

### Permafrost

Permafrost is a crucial component of northern boreal forests. Permafrost contains substantial carbon (~1.3-1.7 teratonnes) and methane (~20 Gt) reserves, stored in frozen organic soils [12-14], far exceeding the carbon stored in the active layer and aboveground biomass [15]. As climate change accelerates, permafrost is at risk of melting, threatening to release ancient reserves in the form of carbon dioxide and methane, and jeopardize ecosystem function. Permafrost thawing and large-scale GHG emissions could further exacerbate climate change, potentially initiating feedback loops [16]. Therefore, high-latitude regions require a management plan to reduce the impacts of melting permafrost on delicate ecosystems. While there is debate about which land covers will best protect ecosystem function, maintain permafrost, and ensure carbon sequestration, there is consensus that action is necessary to help ecosystems adapt to anthropogenic climate change [17].

While an overlap between Canada's boreal treeline and permafrost line may suggest that forests affect permafrost negatively, there is ample contrary evidence that forests help maintain permafrost in many ways [18] (see Box 2a). The results from the experimental station in Farmers Loop (Fairbanks) run by the US Army Corps of Engineers, Cold Regions Research and Engineering Laboratory (CRREL) demonstrate the role forests play in maintaining the stability of permafrost [19] (see Box 2b). Data from other monitoring sites across the world support this conclusion that forest removal results in an increase in active layer thickness and ground temperature [20-22]. These findings are further validated by modeling studies which reveal positive relationships between forest cover and permafrost integrity [23-25]. Even in the larger boreal, average winter soil temperatures are found to be significantly lower in forested sites compared to open lands [26], pointing to forests altering the ground thermal regime favorably.

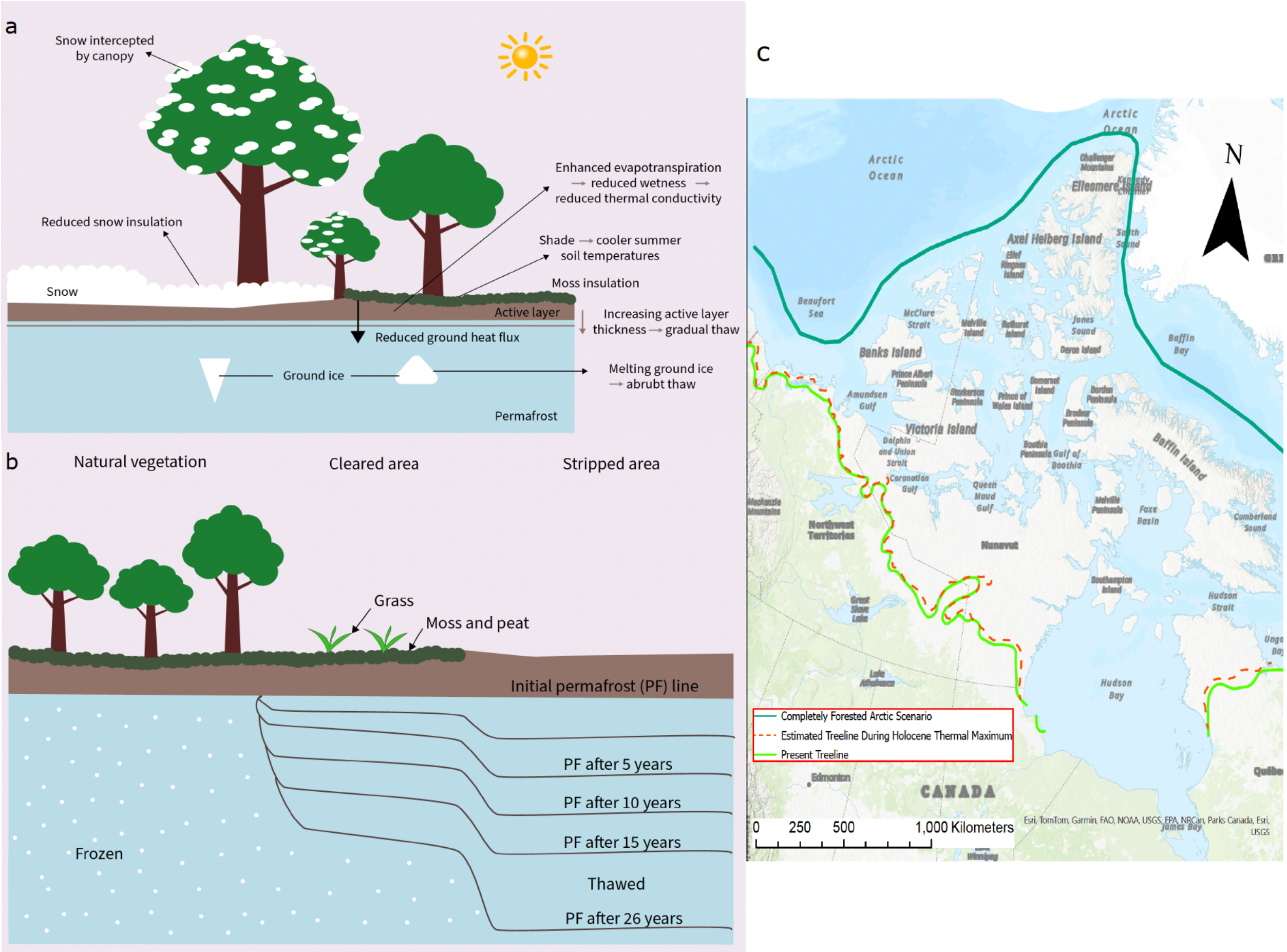

Box 2: a) An illustration showing forests-permafrost dynamics. Forests reduce soil temperatures during summer because of shading and reduce snow insulation during winter due to reduced forest floor accumulation. Enhanced evapotranspiration (ET) in forests reduces soil wetness and therefore the thermal conductivity, preserving permafrost. The interaction between forests and moss layers also play an important role in maintaining permafrost stability. b) The CRREL experimental station in Farmers Loop that monitored different ground covers for 26 years. The site was separated into three segments. A segment where the natural vegetation was untouched (left). A cleared area where trees and major growth was removed, but small shrubs, grass, and moss layers were allowed to grow (middle). A stripped area where all vegetation including moss layers were continuously stripped (right). Permafrost levels were measured regularly over 26 years. Results showed that forests preserve permafrost and any clearing of vegetation significantly exacerbates permafrost melt. c) Tree line extents during present day and altered climate states. A climate state in which the boreal fully expands into the arctic does not reach equilibrium and is pushed back to a climate state with an extended tree line that might have existed during the Holocene thermal maximum (HTM), suggesting that the warming from overall climate feedbacks is insufficient to push the boreal forest from the HTM tree line to complete Arctic forestation.

Forests alter the ground thermal regime, reducing the impact of rising summer air temperatures on soil temperatures [27, 28] (see Box 2a,b). Additionally, the reduced accumulation and prolonged melting of snow on the forest floor, compared to open lands, reduces the extent of snow-trapped insulation during winter (see Box 2a, section "relationship between snow and tree cover") [27, 28]. In spring, the snow albedo effect reduces soil warming by slowing down melting, more so on the forest floor due to radiation interception by the canopy. Moreover, forests reduce ground heat flux by redistributing intercepted energy towards sensible and latent heat fluxes (see Box 2a,b, supplementary Fig. 3). Forests also influence the thermal diffusivity of the soil by creating insulating soil layers and mediating soil moisture [29]. By enhancing evapotranspiration (ET), forests reduce soil wetness, which in turn reduces thermal conductivity [27, 29] (see Box 2a,b, supplementary Fig. 3). Furthermore, mosses, constituting a substantial portion of southern arctic vegetation, form thick insulating mats that shield the soil from warmer surface temperatures [29, 30], highlighting the importance of understanding interactions between forest and moss layers (see Box 2a,b, supplementary Fig. 3). The impact of these vegetation-related effects on the depth of the active layer and various thawing regimes remains unclear. While gradual thaw can increase soil decomposition, releasing nutrients and enhancing vegetation productivity, abrupt thaw (also known as thermokarst) can occur in regions with high ice volume, causing soil collapse and affecting local vegetation growth [27] (see Box 2a, supplementary Fig. 3).

Permafrost is going to respond to climate change, with rising summer temperatures and increased precipitation (see section "changing climatic conditions"). Understanding the response of permafrost to Earth's previous warm periods is crucial to plan for effects of future warming. The mid-Pliocene warm period (mPWP, ~3.264 to 3.025 Ma BP) serves as a valuable analogue for projected climate change scenarios [31]. Research indicates that near-surface permafrost during the mPWP was significantly reduced, estimated to be approximately 93% smaller than pre-industrial levels, coinciding with elevated surface air temperatures and increased winter snow accumulation [31]. This finding indicates that permafrost will thaw significantly as the climate changes in the coming decades, with major impacts on climate, hydrology, and ecosystems [12, 31, 32]. Therefore, the role of forests in regulating permafrost dynamics at high latitudes is crucial and cannot be overlooked in afforestation assessments.

**Observation from Earth's climatic history**

Examining the historical northward expansion of boreal forests and treelines provides valuable insights into positive feedback loops between forests and the climate, as well as crucial corrective mechanisms. Regarding boreal forest expansion, the Sahtu Nation in the Northwest Territories believed that the treeline extended to the Arctic Ocean 9000 years ago, much further north than the present treeline [33]. While there is no consensus on the exact extent of the treeline during the Holocene, it is observed that trees colonized quickly behind retreating glaciers in Canada, and the treeline stabilized thousands of years ago in some areas. For example, the Quebec treeline has remained relatively stable for the past 6000 years, with varying species and temperature gradients throughout the Holocene [34]. This treeline stability supports the argument that the boreal treeline may not continually move north, reinforcing itself, but significantly influences the preferred position of the arctic front [34-37].

Some studies suggest that the mid-Holocene (6 ka BP) high-latitude warming cannot be attributed to orbital forcing alone and require positive feedback from the northward expansion of boreal forests to explain the Holocene thermal maximum (HTM) [38]. Paleobotanical evidence supports the notion that boreal forests indeed migrated northward in response to orbital forcing [38-40] (see Box 2c). Global climate models estimate that this expansion may have contributed an additional 4°C in spring and 1°C in other seasons [38], but studies disagree on the exact contribution of vegetation to this warming [41-45]. Moreover, some studies dispute the role of vegetation feedbacks during the HTM and argue that climate models may have overestimated the positive feedbacks from the expansion of the boreal forest into the tundra [46]. Paleoceanographic observations suggest that parts of the North Atlantic were approximately 4°C warmer than present day during the mid-Holocene. Climate models that incorporate mid-Holocene North Atlantic Sea Surface Temperature (SST) and sea ice conditions estimate that a significant portion of the high-latitude warming can be attributed to SSTs, orbital forcing, and sea ice [47]. The role of vegetation feedback is further explored by studies that investigate possible equilibrium states in the Earth's climate under specific boundary conditions [48, 49]. These studies observe that despite initial forest extension, warming from feedback between ocean, land, atmosphere, and sea ice is insufficient to continually push the boreal forest north into a different equilibrium state [48, 49] (see Box 2c). This suggests that despite feedbacks between climate and land cover at high latitudes, vegetation extent may be stable in response to reasonable perturbations [48].

Regardless of the ongoing debate about the role of positive vegetation feedbacks during the HTM and the extent of the boreal treeline during the Holocene, it is essential to recognize that a warmer and higher $CO_2$ climate state may create unprecedented conditions that have not been seen in Earth's recent geological past, leading to unpredictable responses from vegetation cover. A thorough examination of vegetation feedback during the mPWP may provide additional insights into this phenomenon [50]. On the other hand, it is also crucial to acknowledge that positive feedbacks alone cannot account for the stability of vegetation at high latitudes during the HTM and the pre-industrial Holocene, indicating that corrective mechanisms in the Earth system play a dominant role.

**Non-radiative processes and energy redistribution**

While change in radiative processes like albedo after afforestation has been recently highlighted in afforestation studies [51-55] (though with large uncertainties, see Box 3c), less attention is given to how forests influence non-radiative processes and energy redistribution [56-58]. Non-radiative processes influence the temperature-based BGP effect and its $CO_2$ equivalent ($CO_2$e) contribution [56, 57], which locally dominates in many afforestation scenarios. While carbon sequestration mitigates warming, the reduced albedo (a BGP effect) of forested regions can increase net available radiation, potentially offsetting the cooling effect through BGC processes [51-55]. Land covers vary in their ability to utilize the net available radiation for work including ET, turbulent heat convection, and photosynthesis [56-58] (see Box 3a). This efficiency in energy dissipation, crucial for controlling the surface energy balance, is characterized by an energy redistribution factor [57].

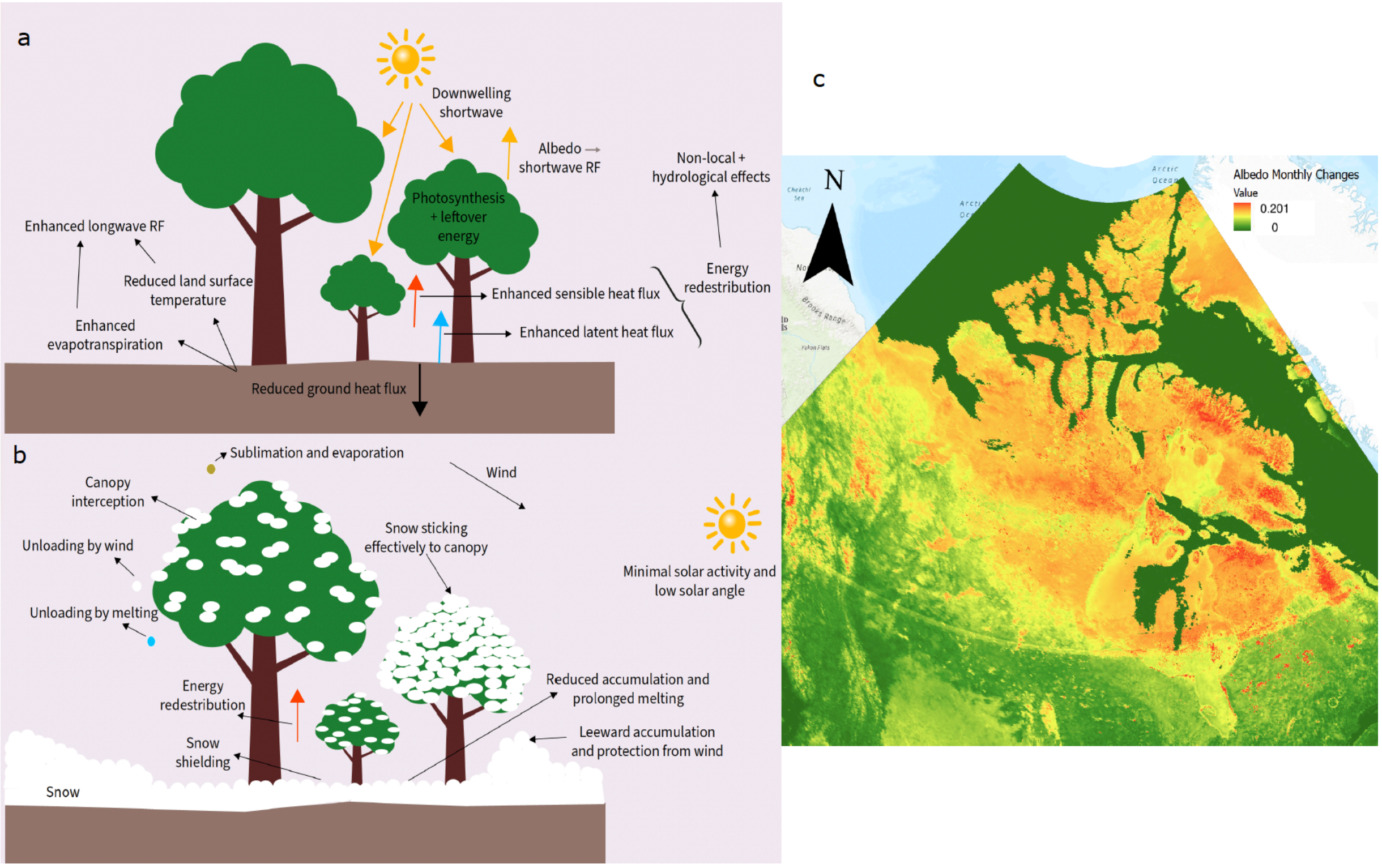

Box 3: a) An illustration showing how forests alter radiative and non-radiative processes. The decreased albedo in forests induces shortwave radiative forcing (RF). However, forests also redistribute the absorbed solar energy into processes such as photosynthesis, latent heat flux, and sensible heat flux. The increased ET from enhanced latent flux decreases surface temperatures, which contributes to local cooling but also induces longwave RF. The effective energy redistribution in forests affects non-local and hydrological processes, modifying atmospheric energy balance. Forests also reduce ground heat flux because of energy redistribution to other fluxes. b) An illustration depicting snow-related processes in forests. Forests with dense canopies contribute to ground snow-shielding, reducing snow-related albedo. On the contrary, both prolonged spring melting and leeward snow accumulation can increase snow-related albedo. Effective canopy interception and resistance of canopies to unloading by melting, wind, sublimation, and evaporation, can increase forest albedo. Minimal solar activity and low illumination angles in the boreal and southern arctic increases adhesion of snow to canopies. c) Average standard deviation of monthly moderate resolution imaging spectroradiometer (MODIS) albedo data. The daily post-processed 500-m global surface blue-sky albedo climatology data is obtained at 0.05° resolution [59]. The daily data is aggregated to monthly standard deviations, and the 12-month average of the monthly standard deviations is computed. Many parts of the boreal have a high standard deviation of 0.1-0.2, because of the distinct snow season and lack of temporal resolution to capture the dynamic interaction between snow and vegetation.

A portion of the net incoming shortwave radiation is photosynthetically active radiation (PAR), some of which is absorbed by trees, with a fraction used for photosynthesis (see section "forest structure and temporal analysis") and the majority converted into sensible or latent heat [60] (see Box 3a, supplementary Fig. 4). The redistribution factor dictates how this heat is distributed, with forests typically exhibiting higher values compared to other land types, indicating more efficient ET and turbulent exchange of sensible heat [57, 61]. Newly formed forests enhance the land's ability to release moisture, cooling the surroundings by altering the surface energy balance from sensible to latent heat [56] (see Box 3a), an effect observed even with small-scale tree cover gain [62]. The extent of this conversion depends on regional humidity, land aridity, and soil moisture levels [63]. Higher moisture content translates to increased sensible to latent heat conversion, also altering cloud cover and precipitation [63-68] (see supplementary Fig. 4, section "alterations in hydrological processes"). Although non-radiative fluxes in forests contribute to local cooling, the resulting lowered land surface temperature (LST) and increased ET generate longwave RFs that can be commensurate with albedo-driven shortwave RFs [64] (see supplementary Fig. 4). Moreover, the dominance of longwave RFs varies spatially, potentially being more pronounced in boreal and arctic regions [69].

Even after considering the merged radiative and non-radiative based $CO_2$e contribution, multiple uncertainties remain, including non-local effects that dominate local ones in climate models, often acting in the opposite direction [70]. Moreover, many BGP effects and their magnitudes depend on afforestation size, including variation in precipitation levels, atmospheric circulation, and cloud cover [57, 67, 70]. These hydrological processes, in turn, affect albedo by altering aridity gradients [63, 65, 66] and radiation balances at the surface [64, 71] (see section "alterations in hydrological processes"). There is also a significant temporal disparity between the processes involved, as forests sequester carbon gradually over many decades, while BGP and hydrological effects manifest in just a few years. These temporal trade-offs are often overlooked in studies, which tend to neglect the yearly variation of gradual processes like afforestation [51, 55] (see section "forest structure and temporal analysis"). Furthermore, afforestation exhibits strong seasonality effects [56-58], with BGP effects being negligible during the boreal summer, but potentially countering BGC benefits during the boreal winter [56, 57, 64]. This seasonality effect poses a dual risk: minimizing cooling benefits during summer when human vulnerability to heat stress is highest, while failing to account for potential adverse impacts of winter warming [56, 57]. Therefore, afforestation interventions must be designed considering non-radiative effects on regional climate, as well as their potential non-local, temporal, and seasonal tradeoffs, as neglecting them can lead to policies detrimental to local climate adaptation and mitigation [57].

**Relationship between snow and tree cover**

Accurately assessing the climate benefits of afforestation requires considering the fine-scale spatial and temporal variations in snow cover, as snow significantly impacts albedo, non-radiative processes, permafrost dynamics, and hydrology. Modeling snow behavior in response to vegetation growth is challenging, and even climate models struggle with snow-related albedo uncertainty at high latitudes [72-77] (see supplementary subsection "reconciliation with climate models"). Studies indicate that the spatial distribution of land cover and vegetation density predominantly influence the snow-albedo feedback in these regions [28]. Investigating snow accumulation on land cover and the mediation of processes such as interception and snowmelt is crucial to understanding the effects of afforestation on snow [78-84] (see supplementary Fig. 5). Observations reveal that open lands generally accumulate more snow than evergreen forests in winter and undergo earlier and faster melting in spring [85] (see Box 3b), but this pattern reverses with reduced canopy density and deciduous forests [86] (see section "forest structure and temporal analysis"). The greatest snow accumulation occurs in openings to the lee of trees, partly due to forests anchoring snow and protecting it from wind erosion and solar radiation [87] (see Box 3b). As a result, snow that would otherwise be blown away is deposited in forested openings, creating zones of retention [88] (see Box 3b). This uneven accumulation and prolonged spring melting due to forests have significant implications for albedo, permafrost thawing (see section "permafrost"), carbon flux, and hydrological cycles.

An important factor modulating forest albedo and energy balance is the interception of snow by forest canopies, followed by melting, unloading or sublimation on the canopy (see Box 3b, supplementary Fig. 5). Canopy height, age, and density control snow accumulation on and beneath the canopy, regulating the energy balance of the forest and thus melting, grain growth, and refreezing at the forest floor [89, 90]. The denser the canopy, the less snow accumulates on the forest floor, and the higher the ground snow shielding, which reduces albedo [79, 89, 91] (see section "forest structure and temporal analysis", supplementary Fig. 5). However, if intercepted snow sticks to the canopy for extended periods, it could increase forest albedo [92]. Snow adheres effectively to canopies in the absence of solar energy, typical of northern boreal edges where winter sunlight is minimal and the solar angle is low [87, 93] (see Box 3b). The canopy also resists snow unloading by wind unless winds are strong and immediately follow the snowstorm [87, 93] (see Box 3b). Therefore, the snow collected on canopies, termed 'Qali' by the Kobuk valley Inuit, may exert the most important control on forest albedo. However, a concerning finding is that although the canopy intercepts a significant percentage of snow, it does not prevent the albedo of the forest from decreasing [94]. Nevertheless, there is little consensus on this matter, and the impact of intercepted snow on albedo at high latitudes requires further investigation [92, 94].

Several local factors, including topography, elevation, slope, and aspect, hinder a global analysis of the impact of forests on snow. Snow interception and accumulation vary significantly with these factors, making region-specific analysis essential. Furthermore, climate change is rapidly altering high-latitude environments, with projected increases in winter temperatures and precipitation over the coming decades. These changes will impact snow interception, accumulation, and melting on afforested land [95] (see supplementary Fig. 5), which must be considered in afforestation assessments.

**Changing climatic conditions**

The Earth's climate is currently undergoing significant changes and will continue to change in the coming decades. Global mean surface temperatures, both over land and oceans, are surpassing previous record highs. A warmer atmosphere can hold more moisture and is expected to alter atmospheric circulation patterns (see Box 4a). Climate change is also impacting snow seasons, altering the composition of tundra biomes, and influencing wildfire and insect disturbances [95-97] (see supplementary Fig. 6). It is crucial to understand how forests respond to this changing climate, as it has significant implications for the productivity of existing forests and new afforestation initiatives [98-100]. In higher latitudes, a warming world is expected to reduce temperature restrictions on vegetation productivity and the duration and extent of snow cover, both of which would decrease the albedo offset [51, 55], and alter non-radiative processes (see supplementary Fig. 6). Moreover, non-radiative mechanisms may dominate in a warmer climate due to their effects on leaf area, canopy conductance, and water vapor [56, 57, 69].

Wildfires are an integral part of boreal forest ecosystems and play a crucial role in the forest carbon cycle. They regulate forests by facilitating forest succession and regeneration, and maintaining plant and animal biodiversity [96, 101]. While humans and lightning strikes initiate roughly equal numbers of fires, most of the area burned in the boreal is due to lightning caused ignitions, and climate change is projected to increase the number of lightning ignitions [96, 101, 102]. Moreover, climate change is predicted to increase various fire-related variables, including frequency of fires, fire season length, severe fire weather, area burned, fire intensity, and emissions [96, 101-105]. Studies suggest that fire occurrence could increase by 75% by 2100 [96] (see Box 4a). While increases in area burned from wildfires are expected to be gradual, threats from population outbreaks and range expansion of endemic forest insect pests are more immediate [106]. Windthrow, the uprooting or breaking of trees due to strong winds and heavy rainfall, is a major cause of tree mortality. Windthrows can significantly alter forest structure, composition, dynamics, and impact both radiation and carbon balance [107, 109], potentially shifting a forest from being a carbon sink to a carbon source [108, 109]. With climate change expected to increase the frequency and intensity of storms, the incidence of windthrows is likely to rise [100]. The effects of climate change on fire, insect, and windthrow regimes have critical implications for afforestation schemes and need to be carefully considered in assessments, particularly because of the potential reversibility of carbon stores in all pools due to these disturbances [110] (see supplementary Fig. 6, Box 4a).

In addition to specific disturbances driving changes in vegetation distribution, a general trend of enhanced vegetation greening is observed at the northern boreal edge and the southern arctic, indicating shifts in recruitment, mortality, and vegetation productivity [111, 112]. These early signs of boreal shift have significant implications for the taiga and tundra ecosystems [113, 114], particularly permafrost thaw, due to altered ground thermal characteristics [27] (see supplementary Fig. 6). Therefore, boreal afforestation assessments need to consider the changing disturbance regimes and the natural progression of vegetation in the boreal, examine the implications of planting more trees in this context, and account for their impact on critical ecosystems like the tundra when in spatial proximity.

a)

| **Climate related variable/event** | **Approximate projected change in 2100** |
|---|---|
| Wildfire area burned | +3.5–5.5 times relative to 1991–2000 |
| Insect disturbances | + but unclear |
| Temperature | +4-5ºC relative to 1961-1990 |
| Precipitation | +14-21% relative to 1961-1990 |
| Permafrost thaw | +16-35% relative to 2000 |
| Humidity | + but unclear |
| Windthrow | + but unclear |
| Cloud cover | + but unclear |
| Solar radiation | Unclear |
| Net primary productivity (NPP) | +50-75% for doubled [$CO_2$] |
| Respiration | + but Unclear |
| Carbon storage | Unclear |

b)

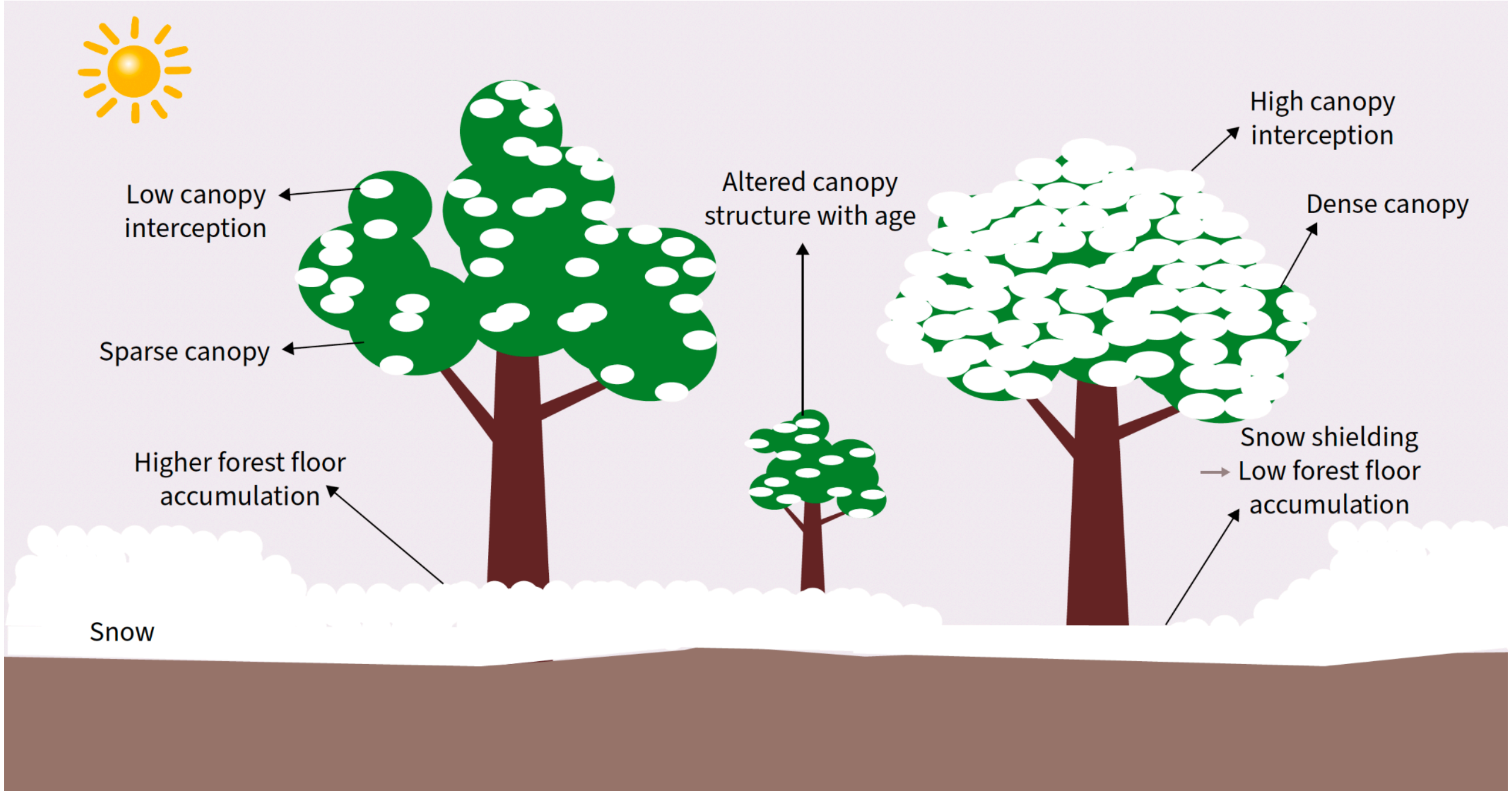

Box 4: a) Projected changes in climate related variables in the boreal [96, 99]. Wildfire occurrence, temperature, precipitation, permafrost thaw, and net primary productivity are projected to increase by significant percentages this century. Insect range expansions, humidity, cloud cover, and ecosystem respiration are estimated to increase, but the exact percentages of increase is unclear. The changes in overall solar radiation and carbon storage is unclear because of the uncertainty in cloud cover for the former, and the uncertainty in the interplay between disturbances, temperature, and precipitation for the latter. b) An illustration showing the difference between how sparse and dense canopies interact with snow. Denser canopies have lower snow accumulation on the forest floor and higher interception at the top of the canopy

(right). Sparse canopies have higher snow accumulation on the forest floor and lower canopy interception (left). The difference in forest structure as trees age also dictates canopy-snow interception and energy redistribution (middle).

**Forest structure and temporal analysis**

Trees absorb photosynthetically active radiation (PAR), which accounts for approximately 50% of incoming shortwave radiation [60, 115]. Only a small portion (around 3%) of this absorbed radiation is utilized for photosynthesis, while the remainder is converted into latent or sensible heat [60, 115]. As trees mature, the net ecosystem productivity (NEP) increases, leading to denser and taller canopies. These canopies, with their intricate leaf structure, absorb more solar radiation [116], resulting in a negative correlation between albedo and canopy density [60] (see supplementary Fig. 7). Studies that map the link between forest structure and albedo globally at high spatial resolution show that forest structure significantly modulates albedo, and is inadequately characterized in existing forest albedo estimation [117, 118]. Moreover, forest structure also plays a crucial role in regulating sensible heat fluxes, which are higher in forests with sparse canopy structures (canopy convector effect) due to low aerodynamic resistance [63, 119] (see supplementary Fig. 7). This canopy cooling through the convector effect suppresses the longwave thermal radiation flux, which the inter-canopy latent heat flux could potentially balance due to the exposed soil surface, but also leads to higher respiration rates and lower NEP [63] (see supplementary Fig. 7). Therefore, uncertainties exist regarding ideal forest structure for climate benefits, and further investigation is warranted.

The remote sensing-based analyses employed by most afforestation assessments substitute space for time and assume instantaneous land cover conversion, overlooking several important details, including: a) the relationship between tree age and canopy structure with albedo [60, 118, 120] (see Box 4b, supplementary Fig. 7), b) The changes in snow interception and unloading with canopy structure and age [91, 117, 121, 122] (see Box 4b, section "relationship between snow and tree cover"), c) the alterations in surface energy redistribution with forest structure and age [120, 123] (see supplementary Fig. 7), and d) the change in canopy density with planting density (see section "effects of planned afforestation projects"). Studies have shown that structural transitions with forest age lead to erroneous albedo estimation (a, b above) due to differences in canopy structure between mature and young forests [117, 120]. Moreover, surface energy redistribution is strongly dependent on forest age [124]. Temporal analysis is crucial, and solely modeling instantaneous conversion for end-of-century responses is inadequate because climate change mitigation policy involves tradeoffs. While maintaining low temperatures by mid-century through cooling measures preserves the short-term climate phase space, $CO_2$ sequestration is essential in the long term. These arguments highlight the importance of integrating forest structure in afforestation assessments, including variation with forest age, plant functional types (PFTs) (species), and planting density, to better capture structural and temporal dynamics [117].

**Short-lived climate forcers**

Land cover changes not only alter BGC processes involving $CO_2$ and water vapor but also impact the concentrations of short-lived climate forcers (SLCFs) including aerosol, ozone, and methane, via the emission of biogenic volatile organic compounds (BVOCs) [125-127]. These emitted BVOCs alter the atmospheric concentration of ozone and methane by reducing the atmosphere's oxidative ability via interaction with other constituents [127, 128] (see supplementary Fig. 8). Furthermore, oxidative byproducts from BVOCs contribute to the formation and expansion of secondary organic aerosol (SOA) particles, which can directly interact with incoming shortwave radiation (direct radiative forcing, DRF) and facilitate the formation of cloud droplets (indirect radiative forcing, IRF) [129, 130] (see supplementary Fig. 8).

Research has demonstrated that forests increase the concentration of SLCFs, with increased ozone and methane contributing to warming and increased aerosols contributing to cooling [125, 127, 129-131]. However, the net RF due to SLCFs from forests is dominated by the DRF and IRF from aerosol cooling, outweighing the warming effects of ozone and methane [125, 132]. Observations reveal that the formation of aerosols and clouds from BVOCs significantly impacts the boreal region, with models underestimating these effects [133, 134]. Therefore, it is crucial to include the RF of SLCFs in afforestation assessments, primarily because the IRF effects from aerosols alone are sufficient to shift forests from being climate-negative to climate-positive [125] (see Table 2 in section "discussion").

**Soil carbon storage and emissions**

It is vital to recognize that the natural climate solutions highlighted by the IPCC include soil carbon sequestration [1], underscored by various land model comparisons [135], SOC measurements in afforested and adjacent areas, and global meta-analyses [136-138]. Therefore, an oversight in many afforestation assessments aiming to identify climate-positive afforestation is the neglect of soil organic carbon (SOC) accumulation over the lifetime of different forest classes, and potential GHG emission reductions due to land-use changes. Forest classes have significantly higher SOC storage advantages than open grasslands, croplands, shrublands, and natural vegetation [139-141]. Moreover, cropland management practices and regular disturbances like tillage worsen soil integrity and enhance organic matter oxidation [142], affecting conclusions regarding afforestation on cropland.

There are many uncertainties surrounding SOC quantification. While promoting carbon sequestration in soils is essential, it may be even more critical to manage soils in a way that prevents permafrost and wetland soils from transitioning from carbon sinks to sources as the climate changes, given that the majority of carbon in the boreal biome is stored in these reservoirs [143]. Research indicates that SOC stocks may initially decrease after afforestation [144], but typically recover over decades [145]. The recovery rate depends on factors such as soil depth [145, 146], climate [146, 147], previous land use [146], and the species planted [148]. Moreover, it is vital to understand how SOC stocks respond to disturbances like wildfires and how long it takes for them to recover, especially as wildfire risk continues to increase in the boreal [149]. The relationship between the size of SOC pool and properties like soil moisture [150], soil Nitrogen concentrations, C:N ratios [151], and pre-afforestation soil carbon [152] is complex and needs more research.

Irrespective of these uncertainties, it is important to include SOC in afforestation assessments. For example, studies considering temperature-based BGP effects observe that including SOC in carbon storage estimates can reduce the net climate-negative regions from approximately 30% to 7% of the total area in high latitudes [56, 57]. This significant reduction highlights the importance of including SOC in assessment frameworks, rather than omitting them to reach an overly simplistic conclusion.

**Alterations in hydrological processes**

Various uncertainties persist regarding the atmospheric adjustments and oceanic feedbacks following afforestation, which may be better captured by effective radiative forcing (ERF) and climate models. Research indicates that instantaneous radiative forcing (RF) overestimates net radiation changes in the boreal region, potentially due to forests' ability to form low-level clouds [60, 70]. These clouds contribute to top-of-atmosphere (TOA) cooling effects, and are also moved non-locally by convection-driven forest breeze [60]. Existing afforestation assessments neglect non-local effects, second-order effects and large-scale climate feedbacks, such as changes in atmospheric circulation patterns (mesoscale circulation, deep convection) and cloud cover formation [56, 67, 70]. Contrary to previous beliefs, these effects are now recognized to be significant even at smaller areal extents of afforestation [70].

Forests are known to enhance ET, which facilitates the formation of shallow cumulus clouds [67] (see supplementary Fig. 9). Research has shown that summertime clouds occur more frequently over forests than over surrounding non-forest regions [67, 68]. Furthermore, observations reveal that clouds tend to form earlier and more rapidly over forested areas, lingering into the evening, possibly due to enhanced thermal flux and atmospheric boundary layer (ABL) moistening [67, 68]. Redistribution of energy, and higher sensible and latent heat fluxes are believed to be key factors driving cloud formation [67]. In addition to driving heat fluxes, forests emit BVOCs that contribute to the generation and growth of SOA particles, thereby facilitating cloud formation [129, 130] (see supplementary Fig. 9). Moreover, clouds play a crucial role in modulating energy balance by altering the quantity of energy reflected, absorbed, and emitted in the atmosphere and at the surface [68, 71]. Thus, clouds influence vertical movements, large-scale circulation, and the hydrological cycle by partitioning energy in the atmosphere [64, 71]. Additionally, clouds mediate outgoing and downwelling shortwave (albedo) and longwave (greenhouse forcing) radiation, controlling the vertical spread of radiative heating. Although the exact impacts clouds have on surface energy balance depend on their altitude, size, and composition, they are known to produce an overall global cooling effect [71].

The impact of afforestation on surface water availability (precipitation minus ET) depends on various factors, including forest and root structure, as well as the precipitation of recycled moisture from afforestation-driven ET, both locally and from upwind locations [65, 66]. While forests generally increase precipitation, they can also reduce rainfall in some regions by decreasing the land surface temperature (LST) and thereby suppressing the thermal contrast with the oceans [65, 66] (see supplementary Fig. 9). Therefore, the impact of forests on the hydrological cycle varies regionally. While altered hydrology such as increased precipitation protects downwind trees from mortality caused by droughts, augmenting climate benefits [51, 65, 66], the effects on surface energy balance are not yet fully understood. For example, both shortwave RF and suppressed longwave RF increase with aridity [63, 64]. While higher net radiation is compensated by increased non-radiative fluxes in these regions, the partitioning of these fluxes also varies with aridity [63]. Sensible heat fluxes are typically higher in drier regions due to the canopy convector effect, whereas latent heat fluxes are higher in humid regions where water is available for ET [63, 64]. Hydrological processes, such as cloud formation, atmospheric circulation, and precipitation, have significant feedbacks on RFs, surface energy balance, and net ecosystem productivity [56, 84, 95]. Therefore, afforestation assessments should make an effort to model some of these feedbacks using Earth system models and reconcile the results with satellite observations to gain a more accurate understanding of the complex interactions involved.

## Methodological limitations

To conduct reliable afforestation assessments, in addition to considering the critical processes discussed in the previous section, it is essential to address methodological limitations. These limitations include uncertainties in remote sensing data and the failure to account for the deliberate and planned nature of afforestation projects, which can impact the accuracy and reliability of the conclusions drawn from afforestation assessments. While remote sensing data is a valuable asset for climate science, enabling the regular tracking of crucial climate variables at global scales, it is important to acknowledge its limitations. For instance, uncertainties in satellite-derived albedo can be as high as 9.7 $W/m^2$ [153], affecting assessments of afforestation that consider albedo [51, 52] (see supplementary section “uncertainties in albedo-related afforestation assessments”). Moreover, the temporal resolution of remote sensing products significantly impacts final conclusions [59, 154] (see supplementary Fig. 1). Remote sensing products are also error-prone in overcast conditions with cloud cover [56, 56], susceptible to bias when the solar zenith angle (SZA) exceeds 70° (particularly relevant at high latitudes during boreal winter) [73], and lack the spatial resolution to account for finer variations in topography [51, 155-157]. Finally, remote sensing land cover products such as the one from moderate resolution imaging spectroradiometer (MODIS) often misclassify land covers, which can significantly bias the final conclusions [51, 52].

Most afforestation assessments use naturally formed forests as a proxy to examine the albedo impacts of afforestation. However, this approach has limitations, as afforestation projects allow for controlled variables such as tree species selection, topography, total afforestation area, and planting density. These factors can be optimized to minimize potential negative impacts. For instance, deciduous trees, with lower albedo offset (higher albedo) than evergreen trees, could be planted in regions where albedo has a significant influence. Topography can be selected to optimize snow cover behavior and illumination angles, mitigating negative BGP effects. The extent of afforestation can be determined by modeling energy balance and hydrological mechanisms to maximize benefits. Additionally, planting density can be adjusted to avoid forest snow shielding issues. Therefore, it is essential to evaluate the climate benefits of afforestation projects on a case-by-case basis, modeling best and worst-case scenarios to account for these factors.

| **Processes/Methods** | **Open Questions/Future Directions** |
|---|---|
| Permafrost | Create physics/data driven landscape-level permafrost maps |
| | Develop a framework using forests to regulate permafrost dynamics and prevent future permafrost thaw |
| | Study post-disturbance soil stability and permafrost thaw and their relationship to forest cover |
| | Investigate how different tree species affect permafrost through variations in canopy structure, root systems, and evapotranspiration rates. |
| | Examine how the combination of canopy cover, understory vegetation, and moss layers collectively influence ground thermal regimes |
| Radiative and Non-radiative Processes | Develop higher resolution (spatial and temporal) albedo land use change maps |
| | Reduce uncertainty reduction in albedo estimates during the snow season |
| | Develop spatial maps of energy redistribution factors and longwave forcings from non-radiative processes |
| | Create landscape-level maps of energy balance factoring in topography, elevation, slope, aspect, SZA, and time of day |
| | Study how the effects of afforestation scale with the size of the afforested area, and at what scales non-local effects become significant |
| | Investigate seasonality of BGP effects and relevance for mitigation and adaptation |
| Forest-Snow | Quantify benefits/drawbacks associated with snow accumulation and melting in forests |
| Interaction | Explore how different forest management strategies affect snow processes and albedo |
| | Study how topography, elevation, slope, and aspect influence snow interception and accumulation in forested areas |
| | Investigate the effects of various climate change scenarios on snow dynamics in afforested areas |
| | Reduce uncertainty in the role of intercepted snow in increasing forest albedo |
| Changing Climate | Quantify climate benefits of forests under projected disturbance regimes (wildfire, rainfall, snowfall, insects, windthrows) |
| | Determine the long-term implications of increased disturbance frequency and severity on the carbon balance of boreal forests, and how this might shift forests from carbon sinks to carbon sources |
| | Investigate how shifts in vegetation distribution due to climate change affect local and global climate feedback mechanisms, such as albedo changes and non-radiative processes |
| | Study how afforestation initiatives can be designed to enhance resilience to climate-induced disturbances and contribute positively to climate mitigation efforts |
| | Determine the realistic potential of boreal afforestation and reforestation in mitigating climate change, considering the risks of carbon reversibility due to disturbances |
| | Investigate current and past vegetation responses to changing climates, and the need for assisted migration |
| Forest Structure | Conduct temporal analysis of climate benefits as a function of species and stand age, focusing on the role of forest structure |
| | Investigate ways in which different canopy structures and ages affect snow interception, unloading, and subsequent albedo changes |
| | Compare the cooling effects of higher albedo in young or sparse forests with the carbon sequestration benefits of mature, denser forests |
| | Study what combinations of canopy density, tree species, and planting densities yield the best balance between carbon sequestration and biophysical climate effects like albedo and sensible heat flux |
| Short-Lived Climate Forcers | Study over what time scales the cooling effects of aerosol-induced DRF and IRF persist, and how they interact with the warming effects of ozone and methane |
| | Investigate why current atmospheric models underestimate aerosol and cloud formation from BVOCs in boreal regions |
| | Study how rising global temperatures affect BVOC emissions from forests, and what feedback effects might this have on climate? |
| | Quantify forest related SLCF effects as a function of species and age |
| Soil Carbon | Reduce uncertainty in SOC accumulation over the lifetime of the forest |
| | Conduct longitudinal studies assessing SOC recovery post-disturbance across different ecosystems |
| | Investigate how increasing temperatures and altered precipitation patterns affect SOC stability and sequestration |
| | Study how the inclusion of SOC alters the net climate impact evaluations of afforestation projects |
| | Conduct comparative studies of SOC changes in afforested areas with different land-use histories |
| | Integrate remote sensing data and machine learning with ground measurements for SOC estimation |
| | Compare soil emissions in non-afforested lands with afforested lands |
| Hydrological Processes | Quantify TOA cooling and energy modulation via clouds formed by forests |
| | Create models that incorporate non-local and second-order effects to better estimate ERF in afforested regions, particularly in boreal zones where RF overestimates net radiation changes |
| | Incorporate afforestation scenarios into Earth system models to simulate potential changes in atmospheric circulation |
| | Conduct studies in regions with varying degrees of aridity to understand how afforestation affects energy flux partitioning and LST |
| | Use climate models to quantify overall hydrological effects of afforested trees |
| | Spatial maps of water availability and relationship with radiative forcing |
| Earth History | Investigate further the positive feedbacks from northward expansion of forests and relevance for a climate with future forcing |
| | Study how boreal forests will respond to future warming and $CO_2$ levels that exceed those of the Holocene, potentially leading to novel climate-vegetation dynamics |
| | Investigate the dominant corrective mechanisms that counteract positive feedbacks to maintain vegetation stability at high latitudes |
| | Analyze paleoenvironmental records from the mPWP to understand vegetation responses under different climate regimes, providing analogs for future conditions |
| Methods | Reduce uncertainty in satellite-derived albedo and remote sensing products under overcast conditions and higher SZA |
| | Design gap-Filling algorithms that can interpolate missing data due to cloud cover or high solar zenith angles, using spatial and temporal patterns from surrounding pixels |
| | Develop machine learning models that can estimate albedo under challenging conditions (e.g., cloud cover, high SZA) using inputs like land cover type, meteorological data, and historical albedo patterns |
| | Conduct landscape-level afforestation assessments factoring in topography, elevation, slope, and aspect |
| | Use deep learning techniques for more accurate land cover classification, reducing misclassification errors between forests, savannas, and other vegetation types |
| | Build models that account for seasonal changes in vegetation (e.g., leaf-on and leaf-off periods) to adjust albedo estimates accordingly |
| | Use LiDAR data to obtain detailed information on forest canopy height, density, and leaf area index (LAI), improving the representation of forests in both remote sensing products and climate models |
| | Design atmospheric correction models that account for aerosol scattering, water vapor absorption, and other atmospheric constituents affecting albedo measurements |
| | Incorporate satellite observations into climate models using data assimilation methods to update model states in real-time, reducing biases |
| | Incorporate models that simulate changes in snow grain size over time, affecting albedo due to metamorphosis processes, and enhance representations of how different forest canopies intercept and retain snow, influencing surface albedo and energy balance. |
| | Use data from satellites equipped with hyperspectral sensors to obtain detailed spectral information, improving material differentiation and albedo estimation |
| | Collect data spanning different seasons to capture the full range of albedo variability due to snow cover and vegetation phenology |
| | Conduct assessments accounting for deliberate nature of afforestation including tree species selection, topography, total afforestation area, and planting density |
| | Design a framework to optimize afforestation decision variables to increase climate benefits (implementing details given in Box 7) |
| Others | Investigate the use of logging to minimize lifetime carbon emissions due to mortality and wildfire in a changing climate |
| | Collect more on-ground data of aboveground, belowground, and soil carbon pools and reconcile with modeling efforts |
| | Reduce uncertainty of various parameters associated with afforestation (see Box 6a) |
| | Investigate mitigation vs adaptation, regional vs global, and near-term vs long-term tradeoffs |
| | Study interaction of climate benefits of forests with biodiversity, economic prosperity, and food, water, and energy security |

Box 5: Some important open questions and directions for future research in the context of afforestation assessments.

## Discussion and the path forward

While carbon sequestration in biomass pools has garnered the most attention in discussions about the climate benefits of afforestation (for a more detailed review and analysis see [4, 158, 159]), numerous questions remain unanswered. Over the long term, the net gain in carbon stocks is determined by the balance between carbon uptake and losses through decomposition and disturbances. One potential way to reduce these losses is through timber harvesting, which could prevent carbon loss due to tree mortality or wildfires. Optimizing both ecosystem storage and storage in harvested wood products (HWP) may offer advantages [160]. However, recent findings suggest that logging may be more emission-intensive than previously thought, potentially turning logged forests into a net source of emissions, even when considering HWPs [161]. Collecting accurate data on carbon pools is critical, and progress has been made in quantifying global carbon storage potential in biomass and soils [8, 9], existing storage in Canada's managed boreal forests [4], and regional afforestation efforts in Canada [162, 163]. Recent modeling efforts have aimed to estimate carbon storage in afforestation pools across the Canadian boreal using spatial reference sites [158]. However, finer spatially explicit modeling and reconciliation with on-the-ground data are needed to improve confidence in estimates and to create more detailed carbon sequestration maps.

Modeling afforestation is a complex challenge, and determining its climate benefits involves a multitude of interlinked processes and regional factors. Research has recently expanded beyond carbon sequestration, acknowledging changes in albedo due to varying tree cover, suggesting that many global biomes may exhibit a significant albedo offset, rendering afforestation climate negative [51-55]. While these studies represent a significant advancement, the form and nature of their conclusions can be misleading when interpreted by the general public and policymakers without sufficient context [164-166] (see supplementary section “uncertainties in albedo-related afforestation assessments”). We acknowledge that it is impossible for any single afforestation assessment to account for all processes and address all methodological limitations. Therefore, we see our work as a synthesis that encourages future research to include more interlinked processes in their modeling, focus on specific regions and their realities, consider practical afforestation scenarios, and acknowledge important methodological limitations. Additionally, we advocate for a separate section that elaborates on whether studies are conclusive enough for regional policy-making and what the general public needs to know. Without such exposition, oversimplified opinions like "trees are bad" may propagate in the public sphere.

The uncertainties and variabilities arising from various non-modeled processes and methodological limitations are significant enough to preclude any definitive conclusions about the climate benefits of afforestation (see Table 2). The variability in monthly aggregated MODIS albedo data exceeds 0.2 in many boreal regions, rendering conclusions from monthly analyses questionable [59]. Topography, a factor entirely ignored by all studies, accounts for around 30% of the variability in surface energy balance [155]. Cloud cover, another overlooked factor, alters RF by ~1.6 $W/m^2$, while overall ERF in climate models has an uncertainty of around 20% [64, 167, 168]. Longwave RF, not included in any existing study, can reach up to 1.1 $W/m^2$ [64]. Non-radiative processes, neglected by most studies, together have a variability of ~10$W/m^2$ [64]. We aim to tackle some of these uncertainties in the boreal and southern arctic regions through modeling studies in future work, with the goal of providing insights for Canadian and global climate policy.

a)

| **Process/parameter/method** | **Approximate associated uncertainty/variability in quantification** |
|---|---|
| Variability in monthly temporal resolution of MODIS albedo | >0.2 [59] |
| Uncertainty of MODIS albedo from overcast conditions | 0.01 [169] |
| Uncertainty of albedo from SZA | 0.05 [73] |
| Variability of energy balance with topography | 30% [155] |
| Uncertainty of energy balance from misidentified or coarse land cover | Unclear |
| Uncertainty of albedo from RF kernels | 15% [51] |
| Uncertainty of ERF | 20% [64] |
| Variability of radiation balance with cloud cover | 1.6 $W/m^2$ [64] |
| Variability of albedo with forest structure | 0.4 $W/m^2$ [117] |
| Albedo bias in climate models | >0.1 [73, 74] |
| Uncertainty of CERES EBAF albedo | 9.7 $W/m^2$ [153] |
| Uncertainty of downwelling shortwave radiation | 10% [64] |
| Variability of radiation balance with precipitation | Unclear |
| Variability of energy balance with SLCFs | 0.12 $W/m^2$ [129, 130] |
| Variability of net positive afforestation area with SOC inclusion | 23% [56, 57] |
| Variability of net positive afforestation area with emission from previous land use | Unclear |
| Uncertainty of albedo from snow related factors | 0.1 [73, 74] |
| Difference in climate sensitivities of $CO_2$ and albedo | 0.5 $W/m^2$ [64] |
| Variability of ET from land use change | 20% [64] |
| Variability of longwave RF from land use change | 1.1 $W/m^2$ [64] |
| Variability of latent heat flux from land use change | 2.5 $W/m^2$ [64] |
| Variability of sensible heat flux from land use change | 8.5 $W/m^2$ [64] |
| Uncertainty of temperature rise from vegetation feedbacks during HTM | 4°C [38, 47] |
| Variability of temperature with permafrost thaw | 12% [96] |
| Variability of permafrost thaw with increased surface energy | 35% [96] |

b)

| | | Temporal-scale | | | | |
|---|---|---|---|---|---|---|
| Spatial-scale | | Seconds-Minutes-Hours | Days-Weeks | Months-Seasons | Years-Decades | Centuries-Millenia |
| | Leaf | Decreased shortwave reflectance decreases albedo | Interception of snow, followed by melting, unloading, or sublimation | | | |
| | | Evapotranspiration affects various forest processes | | | | |
| | Tree | | Control of tree height and age on snow accumulation | Variation in albedo by species type - deciduous have higher albedo than evergreen | Radiative forcing effects of short-lived climate forcers - aerosol cooling dominates ozone and methane warming | |
| | Stand | | Energy redistribution to latent and sensible heat | Stand-level seasonal modification of ground thermal regime in favour of permafrost preservation | Prevention of permafrost thaw | |
| | | | | Variation in albedo and energy redistribution by forest structure - interaction with snow and high uncertanty | Variation in albedo and non-radiative processes as stands age | |
| | | | Lowered local land surface temperature generates longwave radiative forcing | Non-radiative processes affect energy balance and hydrology | soil carbon recovery as a function of soil depth, species, climate, C:N ratios, and previous land use | |
| | | | | Snow accumulation and melting - affecting albedo, energy redistribution, permafrost dynamics, and hydrology | | |
| | Landscape | | | Variation in snow interception and accumulation by topography, elevation, slope, and aspect | Carbon sinks and sources - post disturbance recovery | |
| | | | | Changes in precipitation and water availability | | |
| | | | | Radiative forcing changes through aridity gradients | | |
| | Biome | | | | Climate change alters disturbance regime and climatic variables | Future permafrost thaw risks derived from previous warm periods like mPWP |
| | | | | | Enhanced vegetation greening and shifts in recruitment | Boreal forests may not cause positive feeddbacks resulting in indefinite expansion into the arctic |
| | Non-local | | | Energy redistribution affects non-local hydrology like atmospheric circulation and cloud cover formation | Cumulative radiative and climatic effects | |

Box 6: a) Uncertainties and variabilities in quantification associated with various processes, parameters, and methods. Percentage changes are mentioned without units, unitless parameters (such as albedo) are mentioned without units, and the rest are mentioned with their respective units (such as W/m2). MODIS - moderate resolution imaging spectroradiometer, SZA - solar zenith angle, RF - radiative forcing, ERF - effective radiative forcing, CERES - clouds and the earth's radiant energy system, EBAF - energy balanced and filled, SLCFs - short-lived climate forcers, SOC - soil organic carbon, ET - evapotranspiration, HTM - holocene thermal maximum. Significant variability exists in remote sensing products, energy balance because of topography and cloud cover, and ERF. SLCF's, non-radiative processes, and forest structure also alter overall RF to a large extent. Moreover, the inclusion of SOC increases net climate benefits of afforestation considerably. b) A summary of important processes and findings arranged by spatial and temporal scale, relevant for afforestation assessments. Refer to Box 5 for open questions and future directions related to these and Box 7 for integration of some of these into an afforestation assessment framework.

While modeling and analysis are essential, the boreal and arctic regions face a significant shortage of field measurements. Therefore, in addition to more comprehensive modeling, we hope that future research also addresses this lack of on-ground data. With an increase in data, researchers can attempt to reconcile remote sensing observations with climate models, which is a major bottleneck in the boreal [72-77] (see supplementary subsection "reconciliation with climate models"). Moreover, this improved model-data synergy, with a strong local focus across the boreal and southern arctic, can be highly beneficial for informing policy decisions regarding afforestation and carefully designing these initiatives. Afforestation in the boreal region can help Canada achieve its mitigation goals while providing adaptation benefits; however, further research is necessary before it can be conclusively stated that afforestation will be climate positive, and these climate-positive regions can be identified.

Our review focused primarily on the Canadian boreal, but the arguments presented are widely applicable. For example, permafrost plays a critical role in afforestation efforts across all circumpolar countries. Similarly, the interplay between forests and snow is a crucial consideration in any region with a consistent snow season. Moreover, radiative and non-radiative processes, hydrological cycles, SLCFs, SOC, GHG emissions from land use change, and a changing climate will all be essential factors in determining the climate positivity of forests worldwide, albeit to varying degrees. The key takeaway is that regional realities must be taken into account in afforestation assessments, as local conditions significantly impact the effectiveness of afforestation efforts.

Afforestation decisions involve tradeoffs. For instance, afforestation can contribute to local cooling. However, the local temperature effects due to afforestation might not harmonize with the global response required, as the primary processes dictating energy balance may differ across spatial scales. This may lead to conflicts between regional needs and global goals, which may not always align. Additionally, processes involved operate at different timescales, with albedo, temperature, and hydrology responding quickly to changes, and carbon sequestration taking longer. The full spectrum of spatial and temporal scales can be found in Box 6b. Forests operate across these scales, with leaf-scale to non-local in the spatial domain, and seconds to centuries in the temporal domain, resulting in inevitable tradeoffs. Therefore, future research must identify specific regional versus global tradeoffs and near-term versus long-term tradeoffs and provide a decision-making framework.

In this work, we primarily examined the climate benefits and drawbacks of afforestation. However, it is essential to recognize that forests also impact other vital Sustainable Development Goals (SDGs), including biodiversity, economic prosperity, and food, water, and energy security. While afforestation can lead to enhanced biodiversity, its implementation without local considerations can harm biodiversity, as well as food and water security, depending on existing land use. These risks can be mitigated by considering regional needs and involving local stakeholders in decision-making. Additionally, we have not discussed in detail the interplay between afforestation and the timber and bioenergy industries, which influence economic and energy security. These considerations raise a philosophical question about prioritizing goals, making tradeoffs, and navigating difficult decisions. We aim to address these questions in the Canadian context in future research.

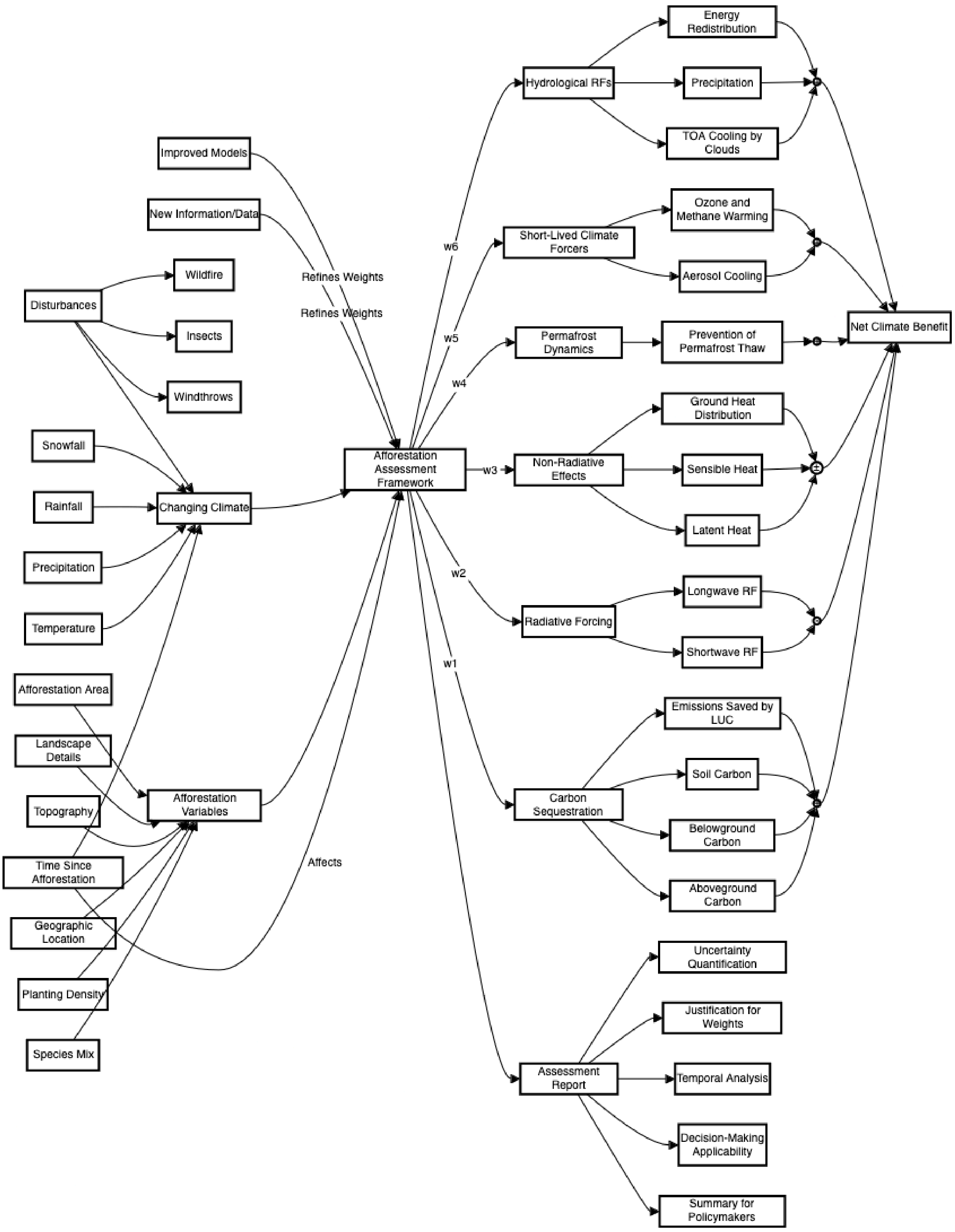

Box 7: Proposed afforestation assessment framework. The framework has six components with their own weights, which are then added to produce a net climate benefit. It also provides an assessment report that includes justification for the weights, results of temporal analysis, and summary for policymakers among other details. Such a framework could be useful to both scientists and policy makers while determining whether afforestation should be carried out in a particular region.

However, to start the conversation around conducting more holistic afforestation assessments for climate benefits, we propose a framework in Box 7. Our proposed assessment framework consists of six components including carbon sequestration, radiative forcing, non-radiative effects, permafrost dynamics, short-lived climate forcers, and hydrological RFs, each with their own sub-components. The effect of each component on the climate is measured in TOA RFs which are weighted, and both the weights and the RFs can vary with time. Therefore, the whole assessment has a temporal component. Moreover, the framework includes a changing climate component which also is a function of time, climate parameters like temperature, rainfall, snowfall, and disturbance regimes such as wildfire, insects, and windthrows. The assessment takes as input afforestation variables including species, mix, planting density, topography, and geographic location, and combines the aforementioned components to produce a net climate benefit. In the end, it provides an assessment report that includes justification for the weights, results of temporal analysis, uncertainty quantification, scope of the assessment in terms of decision making, and summary for policymakers. Determining the appropriate weights for the various components remains an area for further exploration. A potential approach is to base the weights on the uncertainties in the forcing estimates, assigning higher weights to more confident estimates, which avoids treating uncertain and confident estimates equally. Temporal analysis within this framework is essential, as it allows for the tracking of the effects of growth, from seedlings to mature trees, across different components. Additionally, it facilitates the inclusion of climate variables and disturbances that evolve over time.

To demonstrate the value of our proposed framework, we reference studies that account for both carbon sequestration and albedo [51, 55]. These studies conclude that afforestation in the northern boreal region has a negative climate impact, as the reduced albedo of forests fully offsets the carbon benefits, primarily due to the extended snow season. However, our framework reveals that this conclusion cannot be accurately reached by considering only two components while neglecting the other four. In fact, incorporating non-radiative effects and short-lived climate forcers (SLCFs) could reverse these conclusions. Additionally, regional factors like permafrost preservation, a significant climate-positive factor due to the vast carbon reserves, must not be overlooked. Lastly, even within the two components considered, the exclusion of longwave forcing and soil carbon renders these conclusions premature.

Afforestation policies must account for various components to accurately calculate the net climate benefit and ensure that studies incorporate temporal analysis, changing climate conditions, and landscape-level specifics in their evaluations. Policymakers should also mandate that assessments include a "summary for policymakers" to clarify whether the findings are suitable for informing policy at national or global levels. This would help avoid confusion with studies that omit key implementation details and are intended solely for research, not policy guidance. Additionally, policymakers should ensure that assessments consider practical factors like planting density, topography, and the uncertainties associated with their conclusions.

## Conclusion

This review synthesizes existing knowledge on the climate benefits of afforestation, identifying gaps that prevent definitive conclusions about its climate positivity or negativity. With a focus on the Canadian northern boreal and southern arctic regions, which are highly sensitive to climate change and relevant to afforestation initiatives, we discuss regional realities and processes that must be considered in afforestation assessments, including permafrost dynamics, non-radiative processes, aerosol forcing, hydrological processes, and snow cover dynamics. We also highlight methodological shortcomings in existing assessments, including the neglect of SOC and GHG emissions changes, inadequate characterization of forest structure, limitations of remote sensing products, lack of temporal and seasonal analysis, and the failure to account for the planned nature of afforestation. We introduce an assessment framework that combines different components to calculate net climate benefit, while considering temporal analysis, changing climatic conditions, and implementation level parameters. We hope that this synthesis encourages future research to address outlined research gaps and the proposed framework drives forthcoming afforestation assessments in the boreal region to be more holistic. Furthermore, we believe that the research gaps and assessment framework discussed in this review will spur useful discussions to inform and improve Canadian and circumpolar afforestation policy.

## Author contributions

Kevin Bradley Dsouza conducted literature review, wrote the article, and created illustrations. Enoch Ofosu, and Jack Salkeld helped with literature review and data visualization. Richard Boudreault, Juan Moreno-Cruz, and Yuri Leonenko were involved in acquisition of funding, editing the article and overall supervision. All authors contributed to discussion and conceptualization of arguments.

## Competing interests

The authors declare no competing interests

## Supplementary information

Supplementary information contains details about albedo-related afforestation assessments, and additional figures and process diagrams referenced in the main article.

# Supplementary

## Uncertainties in albedo-related afforestation assessments

Research has recently expanded beyond carbon sequestration, acknowledging changes in albedo due to varying tree cover [1-5], suggesting that many global biomes may exhibit a significant albedo offset, rendering afforestation climate negative. Forests modify a land cover's albedo by reflecting less shortwave radiation, retaining more energy on the ground, and majorly influencing surface temperature. The claim that forests are beneficial for carbon sequestration but detrimental to albedo, and therefore climate benefits should be evaluated based on the trade-off between the two, appears straightforward at first. However, this claim is overly simplistic to be useful in any practical setting for several reasons. First, forests do not alter carbon and albedo alone, but are involved in several Earth system processes (permafrost dynamics, non-radiative processes, aerosol forcing, hydrological processes, snow cover dynamics), none of which can be satisfactorily ignored. Second, the methodology employed in most albedo studies has major limitations that can reverse their conclusions (neglect of soil organic carbon (SOC) and greenhouse gas (GHG) emissions change, inadequate characterization of forest structure, limitations of remote sensing products, ignoring temporal and seasonal analysis, not considering planned nature of afforestation). Third, afforestation is a local endeavor influenced by regional realities and landscape-level details, which global remote sensing analyses cannot capture. Fourth, forests will be significantly impacted by the changing climate, and any afforestation analysis that neglects this is incomplete. Moreover, these studies overlook the complex interplay between disturbance-based mortality, albedo, and carbon uptake, which is a crucial consideration in understanding the effectiveness of afforestation efforts. Finally, considering only the positive feedback from albedo, tree growth in the southern arctic would theoretically lead to a runaway forest expansion due to reduced albedo increasing temperatures. However, there is no evidence from Earth's climatic history to support this claim, suggesting that important corrective mechanisms are being ignored in studies that focus solely on albedo.

In addition to the above arguments, the methodological uncertainties in albedo-related afforestation studies pertaining to remote sensing products, climate models, and radiative forcing (RF) kernels are discussed in more detail below.

**Remote sensing products**

It's important to acknowledge the uncertainties in remote sensing products used in albedo-related afforestation studies [1, 3]. For example, uncertainties in satellite-derived albedo can be as high as 9.7 $W/m^2$ [29]. While studies using remote sensing data to investigate the effects of land cover on albedo [1, 3] employ high-spatial-resolution albedo land use maps (LUMs) [6], they ultimately analyze data in monthly aggregates, which can mask important details. Temporal resolution significantly impacts final results [7], and this is particularly evident when examining the variability of albedo climatology (see Fig. 1). In regions with distinct snow and non-snow seasons, such as the boreal and tundra, coarser resolutions may fail to capture dynamic snow cover and clearing on vegetation [7]. Notably, the 8-day aggregate shows an albedo variability of less than 0.1, while the monthly variability exceeds 0.2 in boreal regions [7] (see Fig. 1). This highlights the need to analyze albedo data at finer temporal resolutions than monthly aggregates.

Remote sensing data has limitations, including its inability to accurately estimate surface albedo and temperature under overcast conditions with cloud cover [8, 9]. Additionally, remote sensing albedo products are susceptible to bias when the solar zenith angle (SZA) exceeds 70°, which is particularly relevant at high latitudes during boreal winter [10]. Furthermore, satellite-derived albedo products lack the spatial resolution to account for finer variations in topography, which affects solar radiation intensity, illumination angles, and snow cover [1, 11-13]. To ensure accurate region-specific albedo offset studies, it is essential to address the significant uncertainties related to overcast conditions, SZA, and topography.

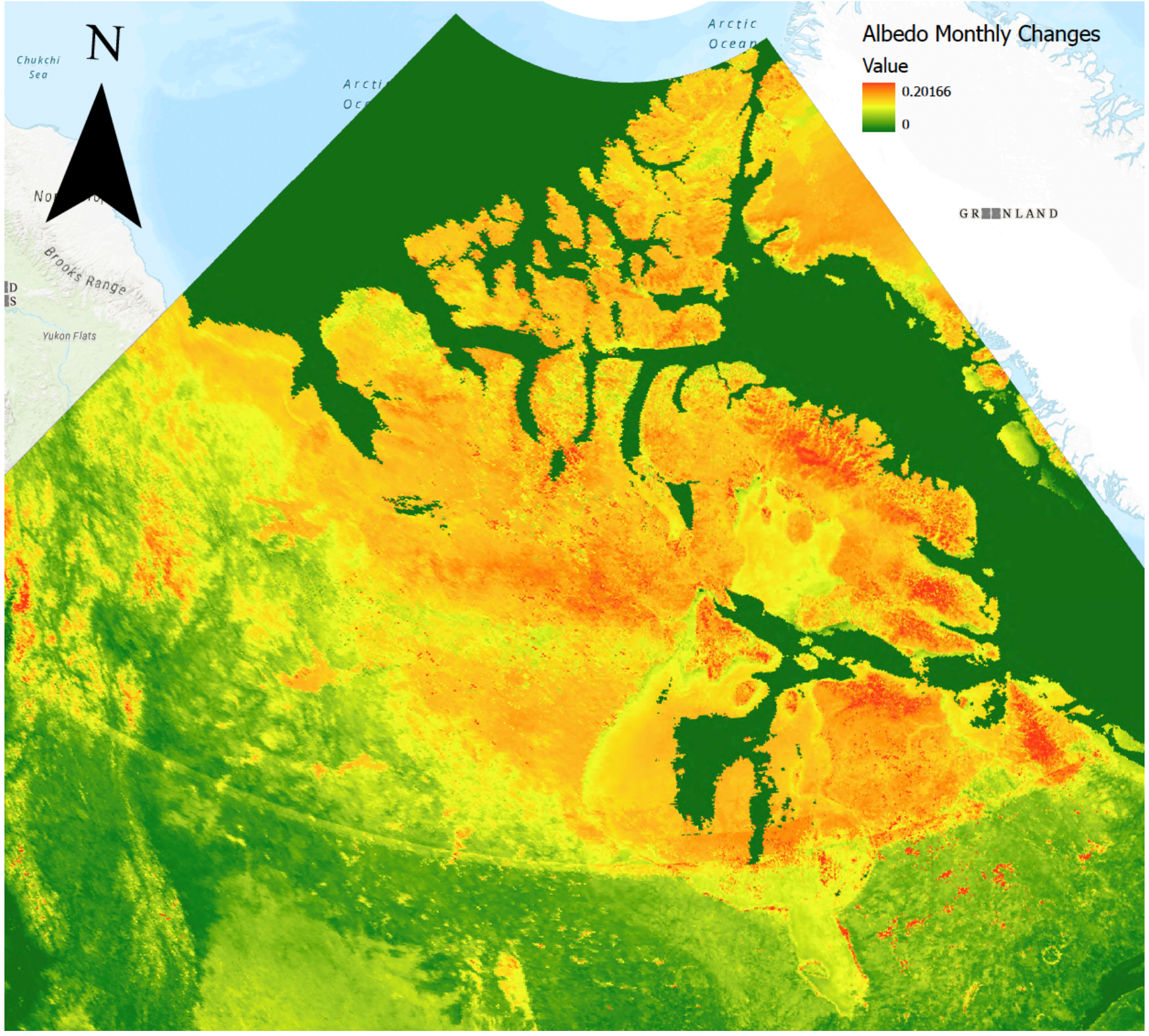

Figure 1: Average standard deviation of monthly moderate resolution imaging spectroradiometer (MODIS) albedo data. The daily post-processed 500-m global surface blue-sky albedo climatology data is obtained at 0.05° resolution [7]. The daily data is aggregated to monthly standard deviations, and the 12-month average of the monthly standard deviations is computed. Many parts of the boreal have a high standard deviation of 0.1-0.2, because of the distinct snow season and lack of temporal resolution to capture the dynamic interaction between snow and vegetation.

Another source of uncertainty arises from land cover classification, both in the original satellite data and post-processing methods. For instance, the monthly moderate resolution imaging spectroradiometer (MODIS) land cover product often misclassifies savannas as forests and patchy forests as savannas [1]. In such cases, studies typically adopt a conservative approach regarding albedo offset, which can bias the final conclusions [1, 3]. Additionally, when sufficient pixel data is lacking, studies often rely on average data for the ecoregion using neighborhood analyses [1]. However, this approach can lead to spatially coarse and biased results. For example, some studies assume the potential end-state land cover for most of the boreal region to be either woody savanna or evergreen needleleaf forests [1] (see Fig. 9 for current land cover [14]). This overlooks the significant presence of deciduous trees, such as trembling aspen, which according to the Canadian National Forest Inventory (NFI) [15] comprise approximately 40% of boreal plains, 13% of taiga plains, and 14% of boreal shield. Given that deciduous forests have a lower albedo offset than evergreen ones due to their structure and dynamics [7], this could lead to a considerable underestimation of albedo values in the boreal region.

**Reconciliation with climate models**

While remote sensing observations offer flexibility in spatio-temporal modeling and global scale analysis, reconciling them with climate model simulations is crucial due to the models' ability to connect observed variables with various Earth system processes, including biogeochemical (BGC), biogeophysical (BGP), and hydrological processes. Discrepancies between remote sensing observations and climate model simulations indicate either incomplete process representations in models or artifacts in observational data. One such discrepancy is the albedo bias in climate models, which can lead to significant uncertainties in snow albedo feedback (SAF), affecting surface energy balance, temperature, and snow behavior [10] (see Fig. 2). This albedo bias exhibits regional patterns, with climate models overestimating observations in boreal regions (positive bias) and underestimating observations in arctic regions (negative bias), likely due to premature snowmelt [16].

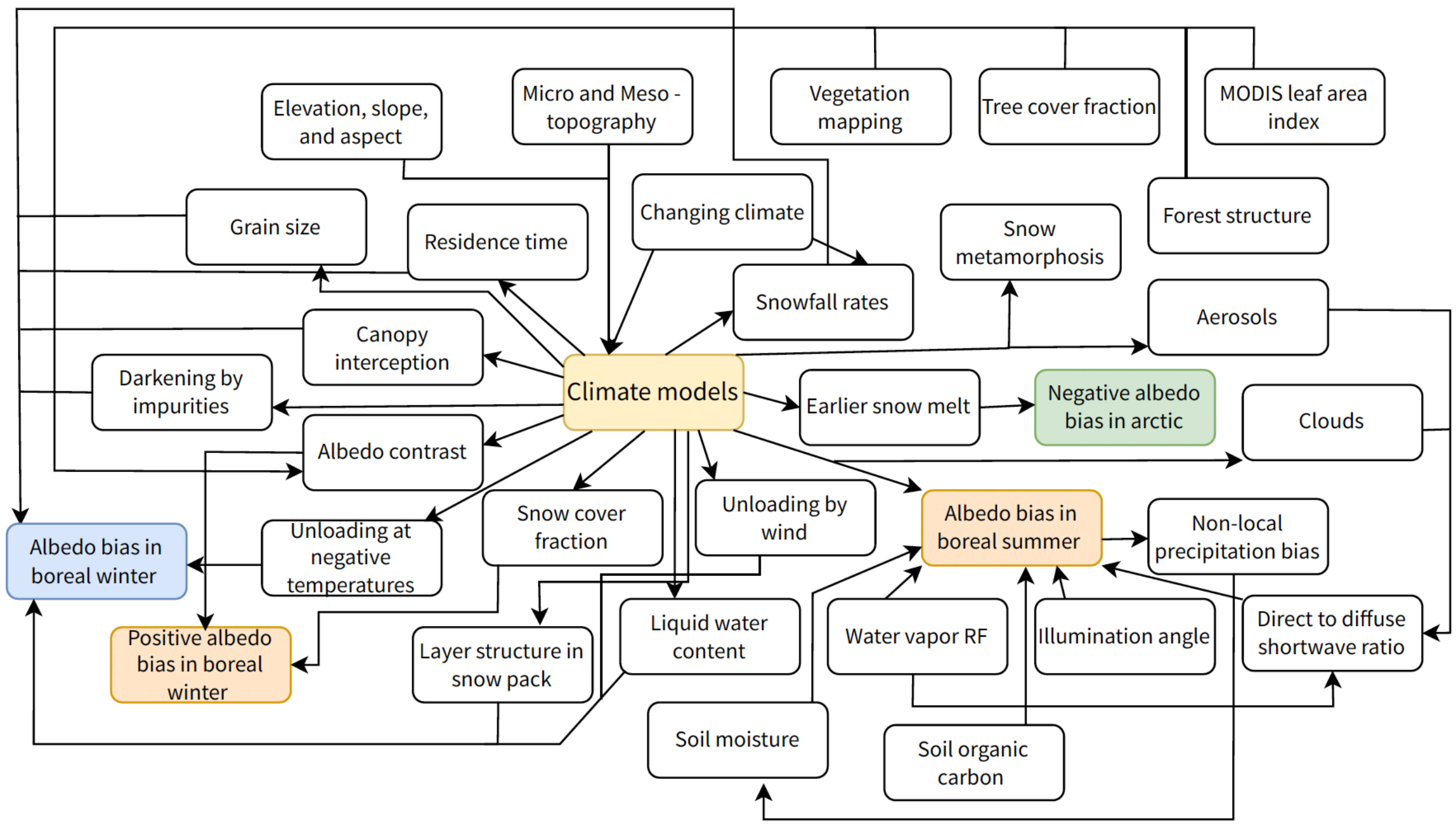

Figure 2: A graph showing the various ways in which albedo bias manifests in climate models. Climate models have insufficient representations of canopy interception, unloading, residence time, darkening by impurities, and snowpack structure and content, which lead to albedo bias in boreal winter. The two major sources of error are the values of snow cover fraction (SCF) and albedo contrast, which contribute to a positive albedo bias in winter. While a modeling of earlier snowmelt might lead to negative albedo bias in the arctic, the bias during the boreal summer could be due to uncertainty in a variety of factors like water vapor RF, soil moisture, SOC, illumination angles, and direct to diffuse shortwave ratio.

Snow is the primary contributor to model biases in boreal regions, due to its high albedo and the lack of consistent measurements for model evaluation. Climate models' representations of snow are associated with multiple sources of uncertainty (see Fig. 2), including: a) snow cover fraction (SCF), b) snow-covered albedo values, c) snow masking by boreal forests, d) canopy snow interception and residence time, e) canopy snow removal at negative temperatures, f) snow unloading by wind at the canopy top, g) effects of snow grain size, h) darkening by snow impurities like black carbon and dust, i) snowpack structure and water content, j) snowfall rates, and k) mechanisms of snow metamorphosis. Research has shown that the multi-model Coupled Model Intercomparison Project Phase 5 (CMIP5) mean albedo significantly differs from the MODIS albedo during boreal winter [10, 17]. These differences can be attributed to SCF errors at lower latitudes and albedo contrast (relative albedo weighting of SCF) at higher latitudes [10, 18]. While models' representations of canopy interception of snow and snow masking effects of forests have been shown to be unrealistic, the impact of other snow-related representations on the albedo bias remains unclear (see e-k above) [17, 19].

Climate models' higher snow-covered albedo values during boreal winter are partly due to inaccurate representations of vegetation mapping, forest structure, tree cover fraction, and leaf area index (LAI) [10, 18] (see Fig. 2). Models with lower LAI values predict higher albedo over boreal forests. However, this is somewhat circular, as climate models calibrate using MODIS-derived LAI, which is unusually low over boreal forests during winter [10]. This discrepancy highlights the issue of uncertain ground truth when field measurements are lacking. Relying solely on MODIS SCF and albedo without reconciling with climate models leaves us vulnerable to other artifacts in MODIS products. While correcting the albedo bias in climate models may seem straightforward, modifying snow-covered albedo values could introduce numerous biases in critical processes like radiative flux, energy flux, and precipitation [10, 19].

Although smaller than the bias during boreal winter, the albedo bias during summer has a greater impact on energy fluxes that influence precipitation, due to higher solar radiation [16, 20]. Research has shown that the correlation between albedo and precipitation bias in CMIP5 models during boreal summer is causal, meaning that albedo bias leads to non-local precipitation bias, not vice versa [18, 20]. For instance, studies have found a negative correlation between terrestrial precipitation and summer albedo, potentially due to either brighter land reducing precipitation or wetter land having lower albedo [20] (see Fig. 2). In addition to snow and vegetation-related factors, the summertime albedo bias may be caused by atmospheric constituents like water vapor that absorb shortwave radiation [18-20]. Other contributing factors could include illumination angle and the ratio of direct and diffuse shortwave radiation, which is affected by scattering from atmospheric constituents like water vapor, aerosols, and clouds [17, 18]. Finally, inconsistent modeling of soil moisture and soil organic carbon (SOC) may also contribute to the albedo bias [20] (see Fig. 2).

Given that albedo interacts with and is impacted by numerous Earth processes, it is crucial to reconcile albedo observations from remote sensing with climate models before drawing definitive conclusions about afforestation based solely on albedo's negative effects. This approach ensures that we do not solely consider the negative effects of albedo in isolation but rather take into account its broader context within the Earth system.

**RF kernels**

The conversion of surface albedo to top-of-atmosphere (TOA) albedo relies on RF kernels, which, despite being used in ensemble predictions, introduce an uncertainty of approximately 15% in TOA conversions [1, 21, 22], which is even more pronounced in the boreal and tundra biomes. Additionally, the relationship between afforestation and clouds is insufficiently characterized in this conversion. Clouds significantly impact the quantification of albedo, as the conversion of surface albedo to TOA albedo relies on vertical profiles of atmospheric optical characteristics, which are influenced by cloud cover [23]. Clouds reduce the surface albedo contribution to TOA albedo through atmospheric attenuation of outgoing radiation, affecting the magnitude of RF associated with afforestation [23]. Furthermore, studies overlook the fact that RF from $CO_2$ and albedo influence different vertical structures, with a similar RF from these two agents resulting in different alterations to surface temperature [23]. To accurately model this effect, it is essential to consider different climate sensitivities before comparing the two.

| Acronym/ Abbreviation | Expansion | Definition |
| --- | --- | --- |
| BGC | Biogeochemical | Related to the transformation and movement of chemical elements and compounds between the Earth, atmosphere, and organisms. Key BGC processes involve elements such as carbon, water, phosphorus, nitrogen, and sulfur, which cycle through ecosystems, sustaining life and regulating Earth's climate and natural systems. |
| BGP | Biogeophysical | The combination of physical, biological, and geological processes acting in a region such as land surface roughness, albedo, and evapotranspiration. |
| GHG | Greenhouse Gas | Atmospheric gases that increase the surface temperature of planets like Earth. What sets them apart from other gases is their ability to absorb the radiation emitted by the planet, which leads to the greenhouse effect. |
| IPCC | Intergovernmental Panel on Climate Change | United Nations intergovernmental body responsible for advancing scientific understanding of climate change resulting from human activities. |
| SLCF | Short-lived Climate Forcer | A group of physically and chemically reactive compounds with atmospheric lifetimes generally less than two decades. This group includes particulate matter (PM), aerosols, and reactive gases like methane, ozone, and volatile organic compounds, among others. |
| SOC | Soil Organic Carbon | This carbon is the primary component of Soil Organic Matter (SOM), which consists of organic residues in different stages of decomposition within the soil. SOC represents the largest carbon reservoir on land, holding more carbon than the combined total in the atmosphere and vegetation. |
| ET | Evapotranspiration | Combined processes through which water moves from the Earth's surface to the atmosphere, encompassing both evaporation of water and plant transpiration. |
| mPWP | Mid-Pliocene Warm Period | A time period within the Pliocene (3.3–3.0 million years ago), when atmospheric $CO_2$ levels were comparable to those of today, and global temperatures were ~3 °C warmer than pre-industrial times. |
| ka BP | Thousand Years Before Present | A time scale primarily used in archaeology, geology, and related scientific fields to indicate when events occurred in |
| | | relation to the advent of practical radiocarbon dating in the 1950s. |
| HTM | Holocene Thermal Maximum | This warm period occurred during the first half of the Holocene epoch, roughly between 9,500 and 5,500 years BP, with the highest temperatures, or thermal maximum, occurring around 8,000 years BP. |
| SST | Sea Surface Temperature | The temperature of the ocean's surface waters, measured using sensors on satellites, ocean reference stations, ships, and buoys. Since the ocean covers a large percent of the Earth's surface, scientists monitor SST to better understand the interactions between the ocean and Earth's atmosphere. |
| RF | Radiative Forcing | It is a concept in climate science used to measure the change in Earth's atmospheric energy balance. This change is influenced by factors such as greenhouse gas concentrations, surface albedo variations, aerosols and shifts in solar irradiance. |
| PAR | Photosynthetically Active Radiation | This is the range of solar radiation between 400 and 700 nanometers that photosynthetic organisms can utilize for the process of photosynthesis. |
| NPP | Net Primary Productivity | This is the amount of carbon generated by primary producers per unit time and area. It is calculated by subtracting plant respiration from total photosynthesis. |
| NEP | Net Ecosystem Productivity | It represents the carbon produced by plants through photosynthesis that is not respired by plants themselves, and heterotrophs. |
| PFT | Plant Functional Type | It is a classification system used by scientists to group plant species based on their similar roles and behaviors within an ecosystem. |
| BVOC | Biogenic Volatile Organic Compound | These compounds produced by plants play essential roles in plant growth. They are crucial in biosphere–atmosphere interactions and significantly influence the chemical and physical properties of the atmosphere. |
| SOA | Secondary Organic Aerosol | Molecules formed through the oxidation of a parent organic molecule over several generations. Unlike primary organic aerosols, which are directly emitted from the biosphere, SOAs are generated either through progressive oxidation of organic compounds or by condensing onto pre-existing particles. |
| ERF | Effective Radiative Forcing | ERF includes both the instantaneous forcing and the subsequent adjustments from the atmosphere and surface. It is a crucial metric for assessing the impact of human activities and natural factors on the climate. |
| TOA | Top-of-Atmosphere | It plays a crucial role in Earth's energy budget, where solar energy enters the Earth system, and reflected light, along with invisible thermal radiation from the Sun-warmed surface, exits. The balance between incoming and outgoing energy at the top of the atmosphere dictates Earth's average temperature. |
| ABL | Atmospheric Boundary Layer | It is the part of the troposphere that is directly affected by Earth's surface and responds to surface influences within an hour or less. |
| LST | Land Surface Temperature | It is a key variable in the Earth's climate system. It reflects processes like the exchange of water and energy between the atmosphere and land surface. |
| SZA | Solar Zenith Angle | It is the angle between the sun's rays and the vertical direction, representing the sun's position relative to the zenith. |
| MODIS | Moderate Resolution Imaging Spectroradiometer | It is an instrument that gathers remotely sensed data, which scientists use to monitor, model, and evaluate the impacts of natural processes and human activities on the Earth's surface. |
| CERES | Clouds and the Earth's Radiant Energy System | They are instruments part of NASA's Earth Observing System (EOS) which are designed to measure both solar-reflected and Earth-emitted radiation, from the top of the atmosphere (TOA) down to Earth's surface. |
| HWP | Harvested Wood Products | They are materials derived from forests, used to create items such as paper, plywood, furniture, or utilized for energy. |
| SDG | Sustainable Development Goals | Adopted by the United Nations in 2015, these goals serve as a universal call to action to protect the planet, and ensure that all people experience peace and prosperity. The 17 Sustainable Development Goals (SDGs) are interconnected, acknowledging that progress in one area impacts outcomes in others, and that development must harmonize economic, environmental, and social sustainability. |

Table 1: Acronyms and abbreviations used in the manuscript along with their expansions, definitions, and importance.

# Figures

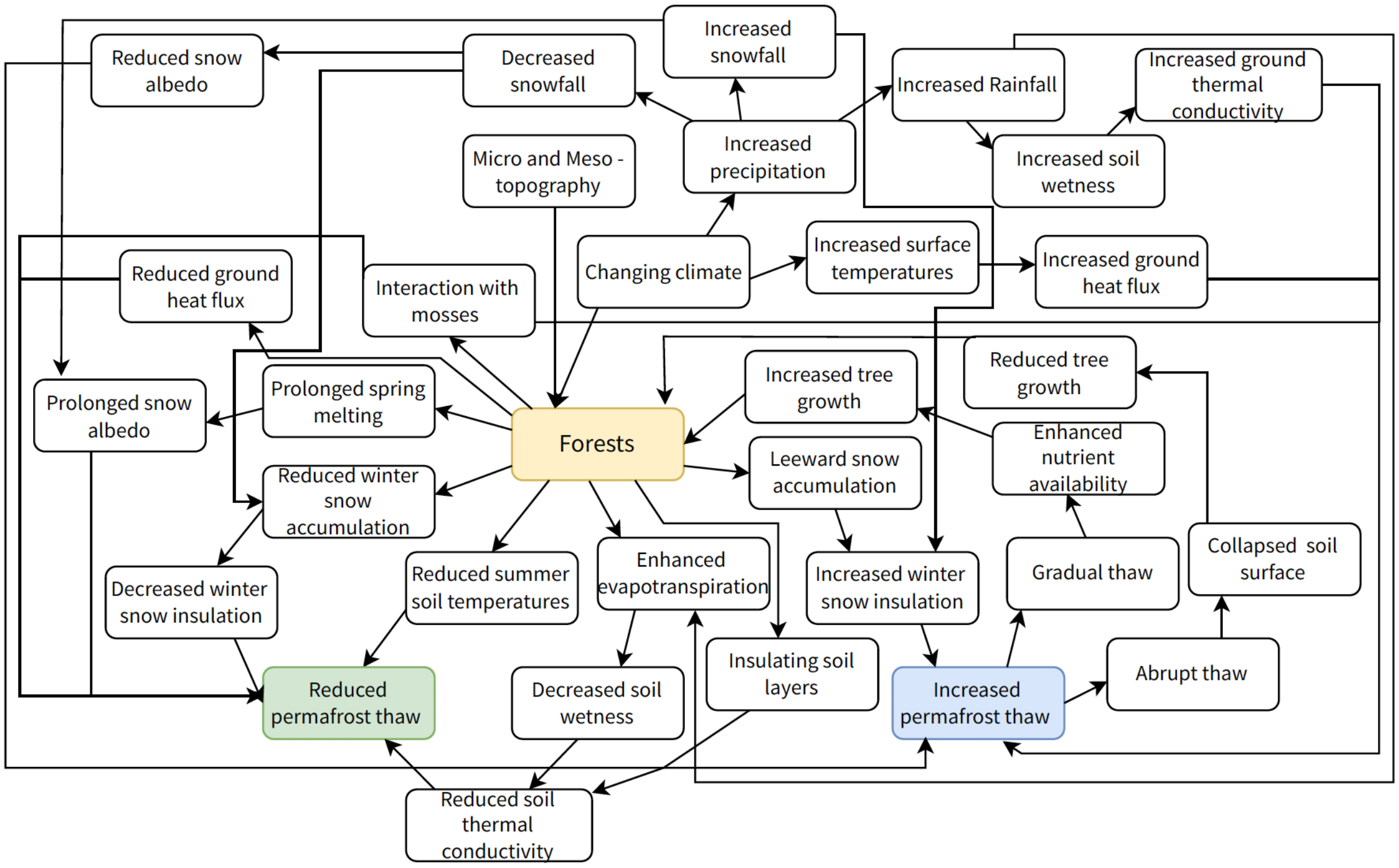

Figure 3: A graph representing the dependencies between forests, permafrost dynamics, and thawing regimes. Forests reduce summer soil temperatures and winter snow accumulation, thereby reducing permafrost thaw. The enhanced evapotranspiration (ET) reduces soil wetness, therefore the soil thermal conductivity, preserving permafrost. The efficient energy redistribution in forests reduces ground heat flux and the prolonged spring melting increases the snow-related albedo. Moreover, the interaction between forests, mosses, and insulating soil layers play a key role in maintaining permafrost stability. Gradual thaw and abrupt thaw have different implications for vegetation productivity, however, both thawing regimes are expected to worsen because of changing climate.

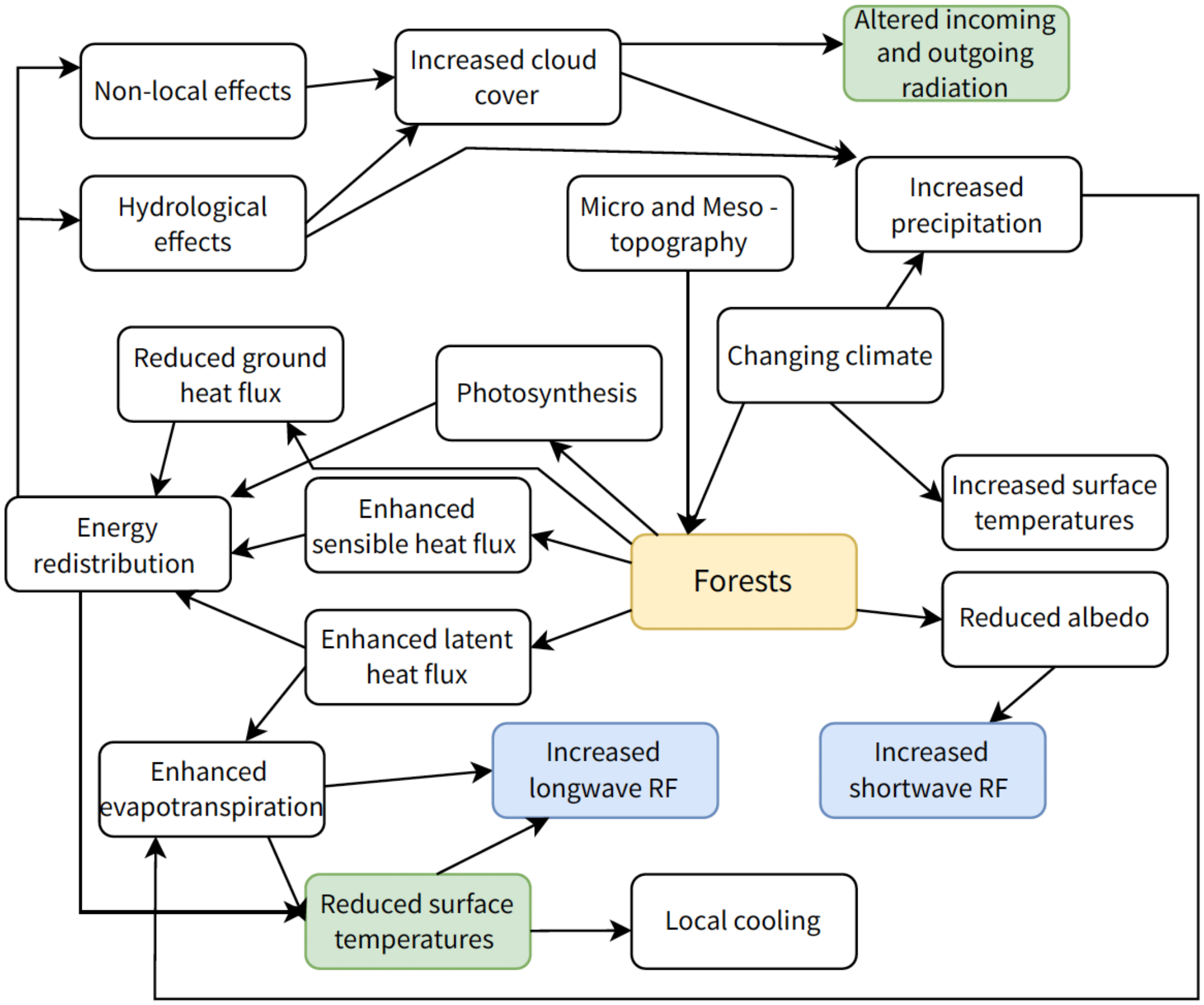

Figure 4: A graph showing how forests affect radiative and non-radiative processes, altering shortwave and longwave RFs, surface temperatures, and net radiation balance. Forests distribute the incoming energy into processes like photosynthesis, sensible heat flux, and latent heat flux. The enhanced ET from increased latent flux reduces surface temperatures. This contributes to local cooling but also induces longwave RF. While the reduced albedo in forests induces shortwave RF, the efficient energy redistribution affects non-local and hydrological processes, which in turn alter atmospheric energy balance. Non-radiative processes are expected to dominate in a changing climate, as temperature and precipitation are key drivers.

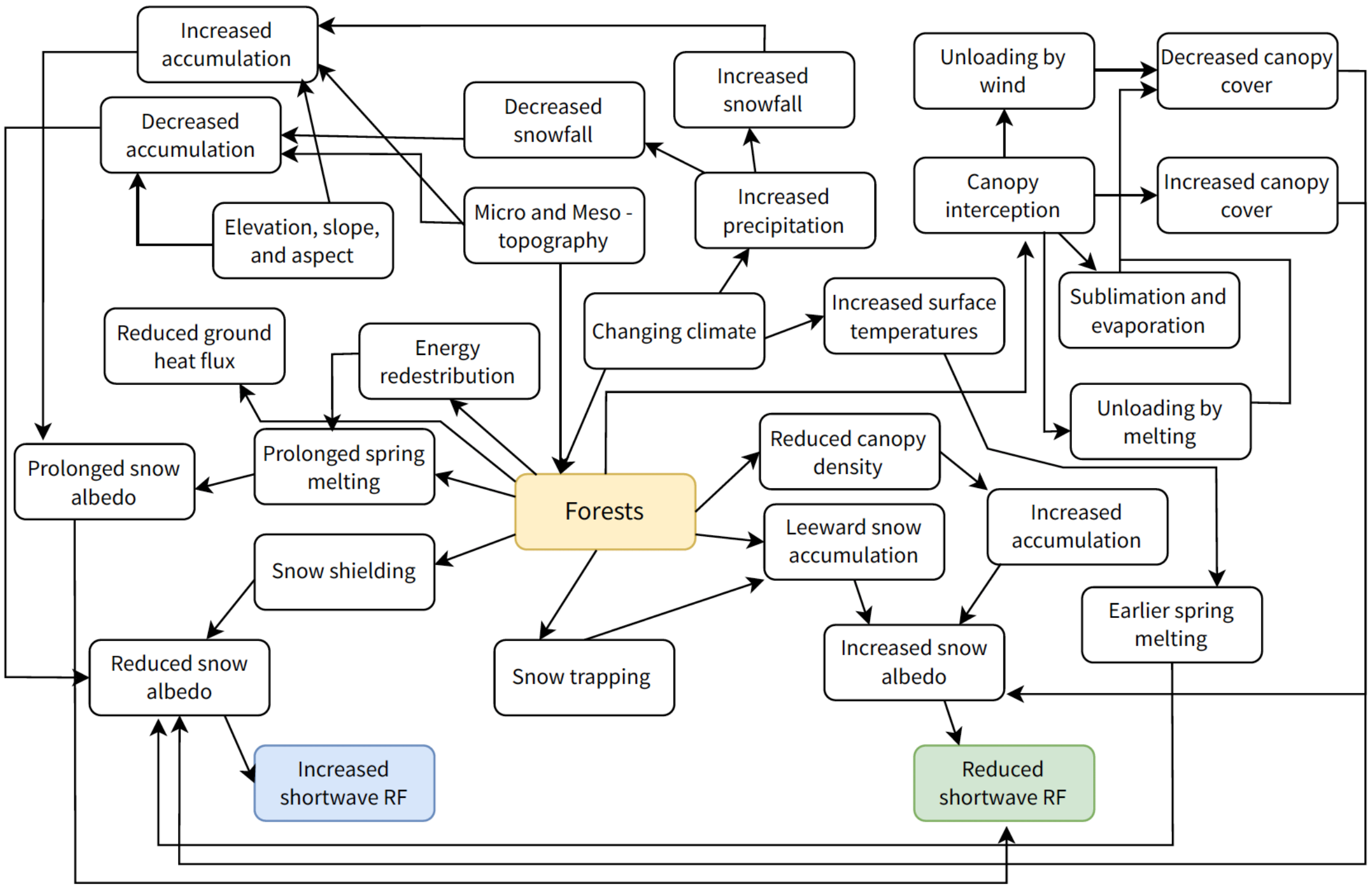

Figure 5: A graph depicting the interaction between forests and snow, changing energy and radiation balance at the surface. Forests with dense canopies shield snow on the ground, thereby reducing snow albedo. On the other hand, both leeward snow accumulation and prolonged spring melting can increase the snow albedo. Forests with reduced canopy density like deciduous forests increase the snow albedo by reducing the snow shielding effect. Efficient canopy interception and resistance to unloading by dense canopies can theoretically increase forest albedo, however, more measurements are needed to confirm this observation. A changing climate will alter the precipitation rates as well as the melting regimes, thereby changing the forest-snow interaction further.

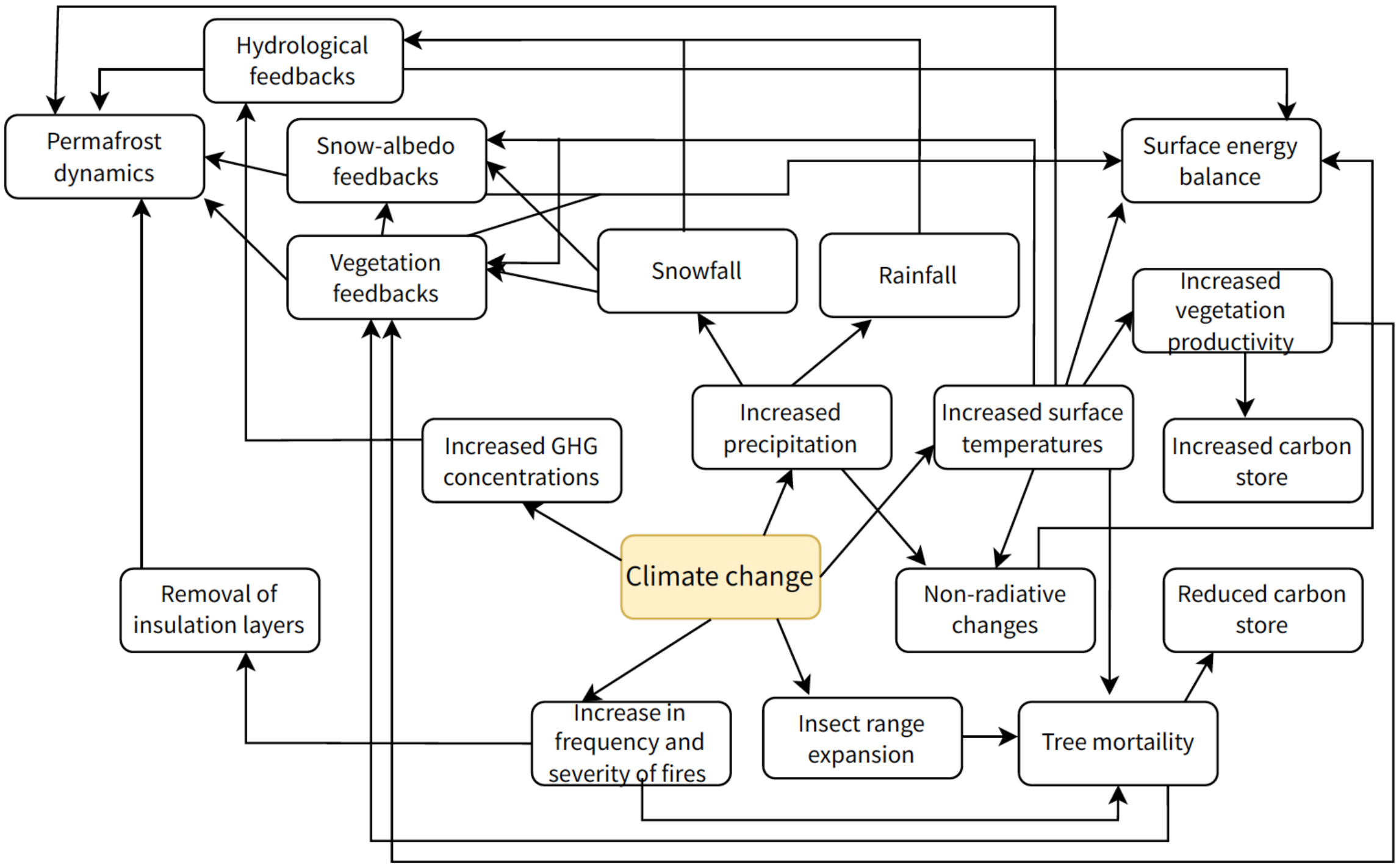

Figure 6: A graph depicting the interactions between forests and a changing climate. Climate change is projected to increase the severity and frequency of forest fires, causing tree mortality and removal of soil insulation layers that protect permafrost. Insect range expansion is also estimated to increase tree mortality and reduce carbon stores. While the increased temperatures might increase vegetation productivity at high latitudes, the increased precipitation is going to alter surface energy balance, and hydrological and albedo feedbacks.

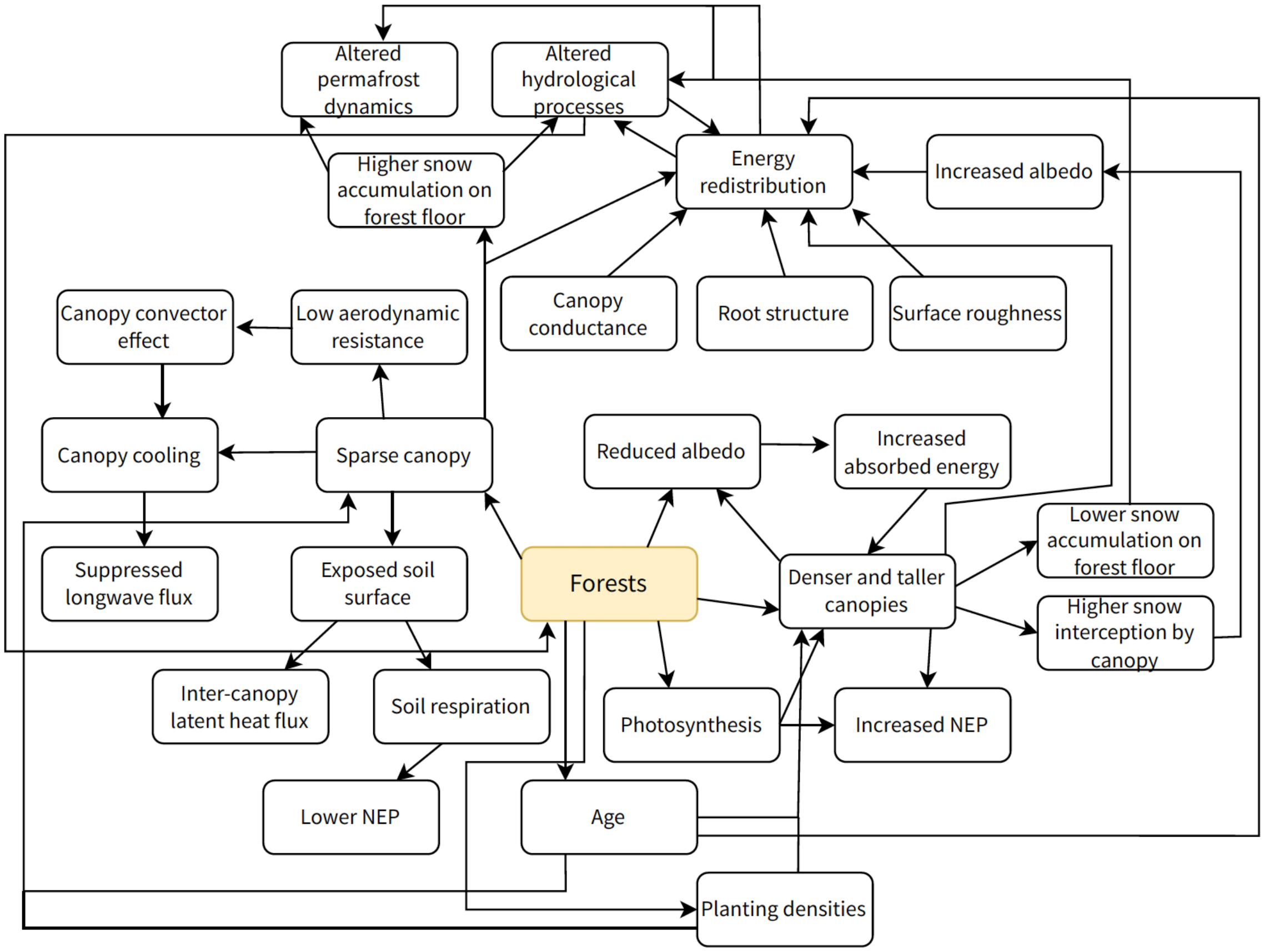

Figure 7: A graph showing the relationship between forest structure and energy balance. The absorbed energy by forests creates denser and taller canopies, which in turn increases absorbed energy. Denser canopies have lower snow accumulation on the forest floor and higher interception at the top of the canopy, both of which affect forest energy balance. Sparse canopies contribute to canopy cooling, but also increase longwave RF and soil respiration. Sparse canopies also have higher snow accumulation on the forest floor, reducing albedo, and altering permafrost and hydrological dynamics. The difference in forest structure as trees age also dictates canopy-snow interception and energy redistribution.

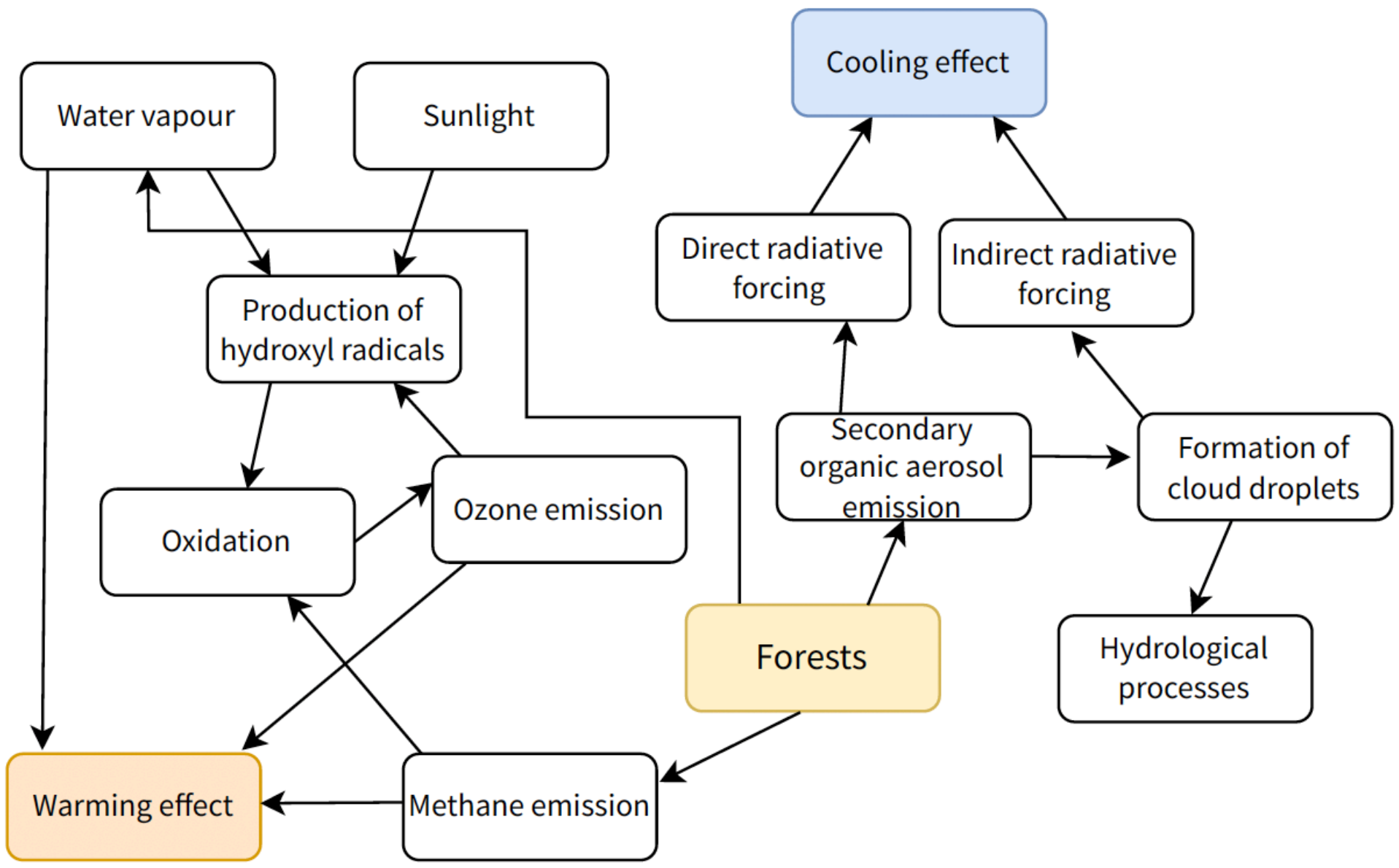

Figure 8: A graph showing the effects associated with BVOC emissions by forests. Forest BVOC emissions alter methane, water vapor, and SOA concentrations. Water vapor produces hydroxyl radicals in the presence of sunlight which react with methane to form ozone. Both methane and ozone cause a warming effect. On the other hand, the SOA particles produce a cooling effect, either through DRF or IRF via cloud formation.

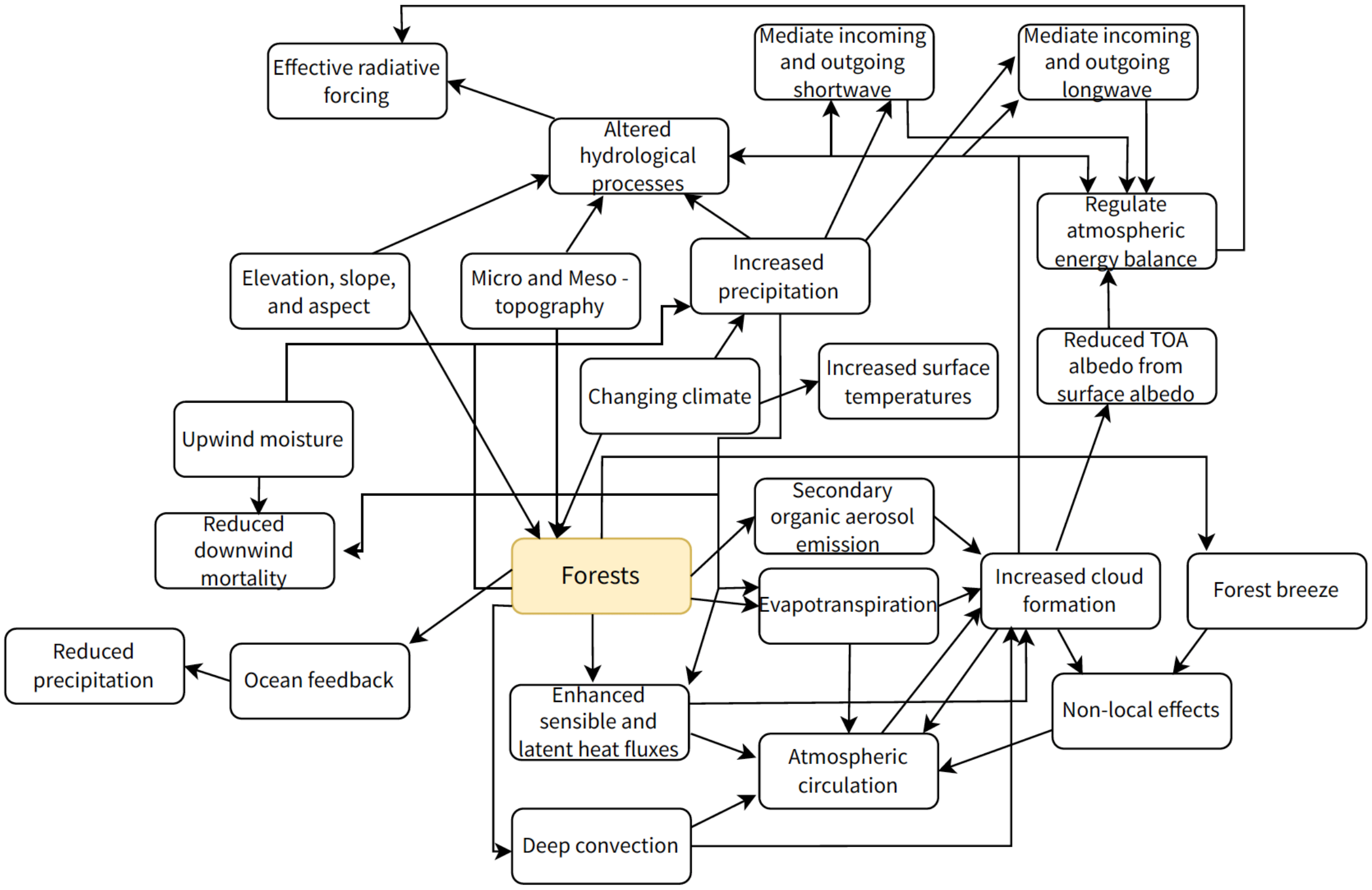

Figure 9: A graph depicting how forests alter hydrological processes. Forests change atmospheric circulation through non-radiative processes, non-local effects, deep convection, and cloud formation. Cloud formation is itself affected by ET and secondary organic aerosol (SOA) emissions from forests. Clouds affect the earth's energy balance by mediating incoming and outgoing shortwave and longwave radiation, contributing to effective radiative forcing (ERF). Moreover, clouds alter the surface albedo contribution to TOA albedo. The increased temperature and precipitation with the changing climate are expected to alter hydrological processes significantly.